\documentclass[suppldata]{interact}

\usepackage{epstopdf}
\usepackage[caption=false]{subfig}

\usepackage[numbers,sort&compress]{natbib}
\bibpunct[, ]{[}{]}{,}{n}{,}{,}
\makeatletter
\def\NAT@def@citea{\def\@citea{\NAT@separator}}
\makeatother

\theoremstyle{plain}

\theoremstyle{definition}

\theoremstyle{remark}

\usepackage[hyphens]{url}
\usepackage{float,epsfig,epstopdf}
\usepackage{graphicx}
\usepackage{amssymb}
\usepackage{amsmath}
\usepackage{bm}
\usepackage[colorlinks,citecolor=blue,linkcolor=blue,urlcolor=blue]{hyperref}
\usepackage{arydshln}
\usepackage{appendix}
\usepackage{pdflscape}
\usepackage{cancel}
\usepackage{multirow}
\newcommand{\bunderline}[1]{\underline{#1\mkern-4mu}\mkern4mu }

\begin{document}

		\title{ Joint modelling of longitudinal measurements and survival times via a multivariate copula approach}
\author	{
		\name{Zili Zhang\thanks{Preprint submitted to J. Applied. Stats}, Christiana Charalambous and Peter Foster}
		\affil{Department of Mathematics, University of Manchester, Manchester M13 9PL, UK}
	}
\maketitle

	\begin{abstract}
	Joint modelling of longitudinal and time-to-event data is usually described by a joint model which uses shared or correlated latent effects to capture associations between the two processes. Under this framework, the joint distribution of the two processes can be derived straightforwardly by assuming conditional independence given the random effects. Alternative approaches to induce interdependency into sub-models have also been considered in the literature and one such approach is using copulas to introduce non-linear correlation between the marginal distributions of the longitudinal and time-to-event processes.  The multivariate Gaussian copula joint model has been proposed in the literature to fit joint data by applying a Monte Carlo expectation-maximisation algorithm. In this paper, we propose an exact likelihood estimation approach to replace the more computationally expensive Monte Carlo expectation-maximisation algorithm and we consider results based on using both the multivariate Gaussian and $t$ copula functions. We also provide a straightforward way to compute dynamic predictions of survival probabilities, showing that our proposed model is comparable in prediction performance to the shared random effects joint model. 
\end{abstract}


\section{Introduction}
In clinical studies, longitudinal data with repeated measurements of response variables at a sequence of informative  (Liang \textit{et al.,} 2009\cite{lia09} and Dai and Pan, 2018\cite{dai18}) or uninformative time points and the time (censored or uncensored) until an event of particular interest occurs (time-to-event data) are often collected jointly for each sample unit. Both of these types of data have been thoroughly studied in their own field and methods have been developed to model and analyse them separately (Verbeke and Molenberghs, 2000\cite{ver00} and  Kalbfleisch  and Prentice, 2002\cite{kal02}). However, separate modelling is inadequate when biomarkers are associated with patients' heathy status, thus correlated with event time. Wulfsohn and Tsiatis (1997)\cite{wul97}  and Guo and Carlin (2004)\cite{guo04} point out a separate or two stage modelling approach (Dafni and Tsiatis, 1998\cite{daf98} and Ye \textit{et al.,} 2008\cite{ye08}) for longitudinal and time-to-event data is likely to produce biased estimation and a joint modelling approach is proposed to alleviate this issue. Exhaustive overviews of joint modelling can be found in Tsiatis and Davidian (2004)\cite{tsi04},  Ibrahim \textit{et al.} (2010)\cite{ibr10},  Papageorgiou \textit{et al.} (2019)\cite{pap19} and Alsefri \textit{et al.} (2020)\cite{als20}. Much of the literature work focuses on linking the two sub-models by either latent classes, shared or correlated random effects (normally a latent Gaussian process). Given latent random variables and covariates of the two processes, conditional independence is assumed between the sub-models as well as within the longitudinal sub-model. There have been a variety of extensions on modelling these unobservable latent variables.  Song, Davidian and Tsiatis (2002)\cite{son02} relaxed the normality assumption of the random effects by allowing them to have a smooth density.  Baghfalaki \textit{et al.} (2017)\cite{bag17} modelled the random effects in the joint model with a finite mixture of multidimensional normal
distributions to propose a heterogeneity joint model (see Verbeke and Molenberghs, 2000\cite{ver00} for homogeneity mixed models). Henderson \textit{et al.} (2000)\cite{hen00}, Wang \textit{et al.} (2001)\cite{wan01} and Xu \textit{et al.} (2001)\cite{xuj01} have discussed applying two correlated elaborate mean-zero stochastic process random effects to characterise correlation between the two sub-models and within the longitudinal measurements. A latent class model for joint analysis of longitudinal biomarker and event process data was proposed by Lin et al. (2002)\cite{lin02}. Liu \textit{et al.} (2015)\cite{liu15} proposed a latent class model with shared random effects, which is essentially a distinct shared random effects joint model of longitudinal and survival data within each latent class.  

The most common and simple shared random effects approach naturally implies the association between the two processes is linear, an assumption which might not always hold. A copula (Hofert \textit{et al,} 2018 \cite{hof18}) is a very useful tool to introduce non-linear correlation between marginals, in our case, the longitudinal and survival processes. The dependency structure is adjustable while keeping the marginal distributions the same. Although the use of copulas to link multivariate or bivariate outcomes either for longitudinal or survival data has been studied extensively, e.g. see Wang (2014)\cite{wan14}, 	Emura \textit{et al.}(2017)\cite{emu17}, K\"{u}r\"{u}m \textit{et al.} (2018a)\cite{kur18a} and (2018b)\cite{kur18b} as well as references therein, not much work has been done on the use of copulas to connect longitudinal and time-to-event data.  Rizopoulos \textit{et al.} (2008a\cite{riz08a} and 2008b\cite{riz08b}) consider applying copulas to specify the joint distribution of the random effects in the two processes instead of assuming common frailty terms. This increases the flexibility in considering different dependence structures between the two sub-models by using various copula functions. Malehi \textit{et al.} (2015)\cite{mal15} adopted the same idea to model random effects but using them to join the longitudinal measurements and time intervals between recurrent events. Alternatively, copulas can link the two marginals directly. Suresh \textit{et al.} (2021a)\cite{sur21a} and (2021b)\cite{sur21b} apply a bivariate Gaussian copula directly on event time and biomarker measured at a single time point and maximise a pseudo-likelihood to perform a dynamic prediction on event time. Diggle \textit{et al.} (2008)\cite{dig08} fitted the joint distribution of longitudinal measurements and the log-transformed event time using a multivariate normal distribution (a special case of applying a Gaussian copula to longitudinal and survival data, where the marginals are normal and log-normal distributions, respectively). This approach is very attractive in terms of computational ease but may lack some flexibility regarding the distribution assumption on event time. Joint modelling of all longitudinal measurements and event time formulated by a multivariate Gaussian copula was proposed by Ganjali and Baghfalaki (2015)\cite{gan15}. The authors use a Monte Carlo expectation-maximisation algorithm for the estimation.  Their approach requires the expectation of the complete data log-likelihood function in the E-step to be calculated under four scenarios. However, some of them cannot be derived analytically thus requiring a Monte Carlo integration, which not only comes with intensive computation but also introduces extra variation/noise into the estimation. The combination with the EM algorithm slows down the computation further and convergence to the maximum likelihood estimators may not always be guaranteed. In this work we improve this estimation process by replacing the Monte Carlo expectation-maximisation algorithm with an exact likelihood estimation approach and also consider using  the multivariate $t$ copula, which allows for tail dependence.  A dynamic method for predicting survival probabilities is also proposed based on the fitted copula joint model.

The remainder of the paper is organised as follows. In Section 2, notations and model specifications are briefly introduced. In Section 3, two simulation studies are conducted for the multivariate Gaussian and $t$ copulas under different dropout (censoring) rates. We firstly investigate the effects of misspecifying the copula used and secondly, we assess the predictive accuracy by computing the mean residual lifetime. In Section 4,  a real data application is carried out based on the AIDS dataset  (Goldman \textit{et al.,} 1996\cite{gol96}) and dynamic predictions of survival probabilities are computed and compared with the predictions under the shared random effects joint model approach. In Section 5, some limitations of this copula joint model are discussed and some future work is proposed.

\section{Copula joint model framework}
Suppose there are $n$ subjects $(i=1,...,n)$ followed over time. Let $\bm y_{i}=(y_{i1},...,y_{im_{i}})$ be the biomarkers measured over time for the $i$th subject. An observed event time $T_{i}=\mbox{min}(C_{i},T_{i}^{*})$ is also recorded for this subject, where $C_{i}$ and $T_{i}^{*}$ denote the right censoring time and true event time, respectively. Let $\delta_{i}=I(T_{i}^{*}<C_{i})$ be the associated censoring indicator, which takes value 1 if the event is observed and 0 otherwise. The time points for the longitudinal measurements and censoring process are assumed to be uninformative.

Consider a longitudinal process with measurement $y_{ij}$ at $t_{ij}$ specified by the following model:
\begin{eqnarray}
	y_{ij}=\bm x_{ij1}^{'}\bm\beta_{1}+\varepsilon_{ij}, \mbox{  }i=1,...,n \mbox{ and } j=1,...,m_{i}\label{longmar}
\end{eqnarray}
where $\bm x_{ij1}^{'},$ the $jth$ row of $\bm x_{i1}=(\bm x_{i11},...,\bm x_{im_{i}1})^{'},$ is a $p$-dimensional vector of explanatory variables with corresponding regression coefficient vector $\bm\beta_{1}$ and  $\varepsilon_{ij}\sim N\left(0,\sigma^{2}\right).$ The single component of $\bm y_{i}=(y_{i1},...,y_{im_{i}})^{'}$ are normally distributed but they are not independent. The within subject correlation for longitudinal measurements $\bm y_{i}$ is introduced by $\bm\varepsilon_{i}=\left(\varepsilon_{i1},...,\varepsilon_{im_{i}}\right).$

A relative risk model with a Weibull baseline function is considered for time-to-event data, given by:
\begin{eqnarray*}
	h_{i}(t)=rt^{r-1}\mbox{exp}(\bm x_{i2}^{'}\bm\beta_{2})
\end{eqnarray*}
where  $\bm x_{i2}$ denotes a $q$-dimensional vector of explanatory variables with corresponding regression coefficient vector $\bm\beta_{2}.$  In other words, the time-to-event data follow a Weibull distribution with shape parameter $r$ and scale parameter $\mbox{exp}\left(-\frac{\bm x_{i2}^{'}\bm\beta}{r}\right).$ A more general but computationally intensive piecewise or B-spline or non-parametric baseline hazard function can also be specified. 

Instead of considering shared or correlated random effects to link the two processes,  the multivariate Gaussian and $t$ copulas are  proposed to join the marginal cdfs of the longitudinal process $(F_{\bm y_{i}})$ and the event time process $(F_{T_{i}}).$  The multivariate Gaussian copula has computational advantages when normality is assumed for the longitudinal measurements as in (\ref{longmar}). The multivariate $t$ copula has more flexibility, as it includes the multivariate Gaussian copula as a special case when its degree of freedom approaches infinity, but this also comes with more computational complexity.  There are some limitations for using other types of copulas due to the typically high dimensionality of the joint distribution of the longitudinal and time to event data. For example, Archimedean copulas, such as Clayton, Gumbel and Frank copulas, with three dimensions or higher are only allowed to have positive correlation between marginals (Yuan, 2007\cite{yan07}).

Let $\bm\theta=\left\{\bm\beta_{1},\bm\beta_{2},\sigma,\bm\rho=(\rho_{y},\rho_{ty}),r\right\}$ denote the vector of parameters for the joint model and 
$\bm U_{i}(\bm\beta_{1},\bm\beta_{2},r,\sigma)=\left(U_{t_{i}}(\bm\beta_{2},r,\sigma), U_{ y_{i1}}(\bm\beta_{1},\sigma),...,U_{ y_{im_{i}}}(\bm\beta_{1},\sigma)\right)^{'}$, where $U_{t_{i}}(\bm\beta_{2},r)=F_{T_{i}^{*}}(t_{i};\bm x_{i2}^{'}\bm\beta_{2},r)$  and $\displaystyle \bm U_{ y_{ij}}(\bm\beta_{1},\sigma)=\Phi\left(\frac{ y_{ij}-\bm x_{ij1}^{'}\bm\beta_{1}}{\sigma}\right)$.  Let $\Phi(\cdot),$ $\phi(\cdot)$ and $\Psi(\cdot;\nu),$ $\psi(\cdot;\nu)$ be the pdf and cdf of standard normal distribution and Student's $t$-distribution with $\nu$ degrees of freedom, respectively.
\subsection{Multivariate Gaussian copula joint model} 
Suppose the multivariate Gaussian copula is applied to characterise the distribution of $\bm U_{i}(\bm\beta_{1},\bm\beta_{2},r,\sigma).$ Let
$\bm Z_{i}(\bm\beta_{1},\bm\beta_{2},r,\sigma)=\left(Z_{t_{i}}(\bm\beta_{2},r),\bm Z_{\bm y_{i}}(\bm\beta_{1},\sigma)^{'}\right)^{'}$, where $Z_{t_{i}}(\bm\beta_{2},r)=\Phi^{-1}\left(U_{t_{i}}(\bm\beta_{2},r)\right)$ is a scalar and $\displaystyle \bm Z_{\bm y_{i}}(\bm\beta_{1},\sigma)=\left(\Phi^{-1}\left(U_{ y_{i1}}(\bm\beta_{1},\sigma)\right),...,\Phi^{-1}\left(U_{ y_{im_{i}}}(\bm\beta_{1},\sigma)\right)\right)^{'}=\frac{\bm y_{i}-\bm x_{i1}\bm\beta_{1}}{\sigma}$ is a vector. The variance-covariance matrix of $\bm Z_{i}(\bm\beta_{1},\bm\beta_{2},r,\sigma)$ can be divided correspondingly, i.e.,
\begin{eqnarray}
	\displaystyle
	\bm R(\bm\rho)_{(t_{i},\bm y_{i})}=\left(
	\begin{array}{ll}
		\displaystyle
		1& \bm R(\rho_{ty})_{(t_{i})(\bm y_{i})}
		\\
		\displaystyle
		\bm R(\rho_{ty})_{(\bm y_{i})(t_{i})}& \bm R(\rho_{y})_{(\bm y_{i})}
		\label{RG}
	\end{array}
	\right).
\end{eqnarray}

Let $\bm\phi_{j}(\cdot;\bm \mu, \bm\Sigma)$ and $\bm\Phi_{j}(\cdot;\bm \mu, \bm\Sigma)$ denote the pdf and cdf of a $j$-dimensional normal random variable with mean $\bm\mu$ and variance-covariance matrix $\bm\Sigma$. When the mean is $\bm 0,$ they are further simplified as $\bm\phi_{j}(\cdot;\bm\Sigma)$ and $\bm\Phi_{j}(\cdot;\bm\Sigma)$.
A subject $i$ with $m_{i}$ number of measurements and $\delta_{i}=1$ has a joint cdf of $(T_{i}^{*},\bm y_{i})$ given by:
\begin{eqnarray*}
	\displaystyle
	F_{T_{i}^{*},\bm y_{i}}(t_{i},\bm y_{i})=\bm\Phi_{m_{i}+1}\left(\bm Z_{i}(\bm\beta_{1},\bm\beta_{2},r,\sigma);\bm R(\bm\rho)_{(t_{i},\bm y_{i})}\right).
\end{eqnarray*}
The corresponding joint pdf is given by:
\begin{eqnarray}
	\displaystyle
	f_{T_{i}^{*},\bm y_{i}}(t_{i},\bm y_{i})
	&=&\sigma^{-m_{i}}\bm\phi_{m_{i}+1}\left(\bm Z_{i}(\bm\beta_{1},\bm\beta_{2},r,\sigma);\bm \Sigma(\bm\rho)_{(t_{i},\bm y_{i})}\right)\frac{f_{T_{i}^{*}}(t_{i};\bm x_{i2}^{'}\bm\beta_{2},r)}{\phi\left(Z_{t_{i}}(\bm\beta_{2},r)\right)}.\label{Gaulik1}
\end{eqnarray}
If this individual is censored at $T_{i}=t_{i},$ the joint pdf is given by:
\begin{eqnarray}
	\displaystyle
	\nonumber	f_{T_{i}^{*},\bm y_{i}}(T_{i}^{*}>t_{i},\bm y_{i})&=&\int_{t_{i}}^{\infty}\sigma^{-m_{i}}\bm\phi_{m_{i}+1}\left(\bm Z_{i}(\bm\beta_{1},\bm\beta_{2},r,\sigma);\bm R(\bm\rho)_{(t_{i},\bm y_{i})}\right)\frac{f_{T_{i}^{*}}(u;\bm x_{i2}^{'}\bm\beta_{2},r)}{\phi\left(Z_{u}(\bm\beta_{2},r)\right)}\,du\\
	\nonumber	\displaystyle
	&=&\sigma^{-m_{i}}\bm\phi_{m_{i}}\left(\bm Z_{y_{i}}(\bm\beta_{1},\sigma);\bm R(\bm\rho)_{(\bm y_{i})}\right)\int_{t_{i}}^{\infty}\phi_{1}\left(Z_{u}(\bm\beta_{2},r);\mu_{i}^{t_{i}|\bm y_{i}},\left(\sigma_{i}^{t_{i}|\bm y_{i}}\right)^2\right)\\
	\nonumber	\displaystyle
	&&\frac{f_{T_{i}^{*}}(u;\bm x_{i2}^{'}\bm\beta_{2},r)}{\phi\left(Z_{u}(\bm\beta_{2},r)\right)}\,du
	\\
	\nonumber	\displaystyle
	&=&\sigma^{-m_{i}}\bm\phi_{m_{i}}\left(\bm Z_{y_{i}}(\bm\beta_{1},\sigma);\bm R(\bm\rho)_{(\bm y_{i})}\right)\int_{Z_{t_{i}}(\bm\beta_{2},r)}^{\infty}\phi_{1}\left(z;\mu_{i}^{t_{i}|\bm y_{i}},\left(\sigma_{i}^{t_{i}|\bm y_{i}}\right)^2\right)\,dz
	\\
	&=&\sigma^{-m_{i}}\bm\phi_{m_{i}}\left(\bm Z_{y_{i}}(\bm\beta_{1},\sigma);\bm R(\bm\rho)_{(\bm y_{i})}\right)\Phi\left(-\frac{Z_{t_{i}}(\bm\beta_{2},r)-\mu_{i}^{t_{i}|\bm y_{i}}}{\sigma_{i}^{t_{i}|\bm y_{i}}}\right),\label{Gaulik2}
\end{eqnarray}
where \[\mu_{i}^{t_{i}|\bm y_{i}}=\bm R(\rho_{ty})_{(t_{i})(\bm y_{i})}\bm R(\rho_{y})_{(\bm y_{i})}^{-1}\bm Z_{y_{i}}(\bm\beta_{1},\sigma)\] and \[\left(\sigma_{i}^{t_{i}|\bm y_{i}}\right)^{2}=1-\bm R(\rho_{ty})_{(t_{i})(\bm y_{i})}\bm R(\rho_{y})_{(\bm y_{i})}^{-1}\bm R(\rho_{ty})_{(\bm y_{i})(t_{i})}.\]

	\subsection{Multivariate $t$-copula joint model}
	Suppose the multivariate $t$-copula with $\nu$ degrees of freedom is applied to characterise the distribution of $\bm U_{i}(\bm\beta_{1},\bm\beta_{2},r,\sigma).$ Let
	$\bm W_{i}^{\nu}(\bm\beta_{1},\bm\beta_{2},r,\sigma)=\left(W_{t_{i}}^{\nu}(\bm\beta_{2},r),\bm W_{\bm y_{i}}^{\nu}(\bm\beta_{1},\sigma)^{'}\right)^{'}$, where $W_{t_{i}}^{\nu}(\bm\beta_{2},r)=\Psi^{-1}\left(U_{t_{i}}(\bm\beta_{2},r);\nu\right)$ is a scalar and $\displaystyle \bm W_{\bm y_{i}}^{\nu}(\bm\beta_{1},\sigma)=\left(\Psi^{-1}\left(U_{ y_{i1}}(\bm\beta_{1},\sigma);\nu\right),...,\Psi^{-1}\left(U_{ y_{im_{i}}}(\bm\beta_{1},\sigma);\nu\right)\right)^{'}$ is a vector. The variance-covariance matrix of $\bm W_{i}^{\nu}(\bm\beta_{1},\bm\beta_{2},r,\sigma)$ can be divided correspondingly, as
	\begin{eqnarray*}
		\displaystyle
		\bm\Sigma(\bm\rho)_{(t_{i},\bm y_{i})}=\frac{\nu}{\nu-2}\bm R(\bm\rho)_{(t_{i},\bm y_{i})}=\frac{\nu}{\nu-2}\left(
		\begin{array}{ll}
			\displaystyle
			1& \bm R(\rho_{ty})_{(t_{i})(\bm y_{i})}
			\\
			\displaystyle
			\bm R(\rho_{ty})_{(\bm y_{i})(t_{i})}& \bm R(\rho_{y})_{(\bm y_{i})}
		\end{array}
		\right),\mbox{ for } \nu>2.
	\end{eqnarray*}
	
	Let $\bm\psi_{j}(\cdot;\nu,\bm \mu, \bm\Sigma)$ and $\bm\Psi_{j}(\cdot;\nu,\bm \mu, \bm\Sigma)$ denote the pdf and cdf of a $j$-dimensional $t$ distribution with $\nu$ degrees of freedom, mean $\bm\mu$ and variance-covariance matrix $\displaystyle\frac{\nu}{\nu-2}\bm\Sigma$. When the mean is $\bm 0,$ they are further simplified as $\bm\psi_{j}(\cdot;\nu,\bm\Sigma)$ and $\bm\Psi_{j}(\cdot;\nu,\bm\Sigma)$. A subject $i$ with $m_{i}$ number of measurements and $\delta_{i}=1$ has a joint cdf of $(T_{i}^{*},\bm y_{i})$ given by:
	\begin{eqnarray*}
		\displaystyle
		F_{T_{i}^{*},\bm y_{i}}(t_{i},\bm y_{i})=\bm\Psi_{m_{i}+1}\left(\bm W_{i}^{\nu}(\bm\beta_{1},\bm\beta_{2},r,\sigma);\nu,\bm R(\bm\rho)_{(t_{i},\bm y_{i})}\right).
	\end{eqnarray*}
	The corresponding joint pdf is given by:
	\begin{eqnarray}
		\displaystyle
		\nonumber f_{T_{i}^{*},\bm y_{i}}(t_{i},\bm y_{i})&=&\sigma^{-m_{i}}\bm\psi_{m_{i}+1}\left(\bm W_{i}^{\nu}(\bm\beta_{1},\bm\beta_{2},r,\sigma);\nu,\bm R(\bm\rho)_{(t_{i},\bm y_{i})}\right)\\
		&\times&\frac{f_{T_{i}^{*}}(t_{i};\bm x_{i2}^{'}\bm\beta_{2},r)}{\psi\left(W_{t_{i}}^{\nu}(\bm\beta_{2},r);\nu\right)}\prod_{j=1}^{m_{i}}\frac{\phi\left(\frac{ y_{ij}-\bm x_{ij1}^{'}\bm\beta_{1}}{\sigma}\right)}{\psi\left(W_{y_{ij}}^{\nu}(\bm\beta_{1},\sigma);\nu\right)}.\label{tlik1}
	\end{eqnarray}
	If this individual is censored at $T_{i}=t_{i},$ the joint pdf is given by:
	\begin{eqnarray*}
		\displaystyle
		\nonumber	f_{T_{i}^{*},\bm y_{i}}(T_{i}^{*}>t_{i},\bm y_{i})&=&\int_{t_{i}}^{\infty}\sigma^{-m_{i}}\bm\psi_{m_{i}+1}\left(\bm W_{i}^{\nu}(\bm\beta_{1},\bm\beta_{2},r,\sigma);\nu,\bm R(\bm\rho)_{(t_{i},\bm y_{i})}\right)\\
		\nonumber	&\times&\frac{f_{T_{i}^{*}}(u;\bm x_{i2}^{'}\bm\beta_{2},r)}{\psi\left(W_{u}^{\nu}(\bm\beta_{2},r);\nu\right)}\prod_{j=1}^{m_{i}}\frac{\phi\left(\frac{ y_{ij}-\bm x_{ij1}^{'}\bm\beta_{1}}{\sigma}\right)}{\psi\left(W_{y_{ij}}^{\nu}(\bm\beta_{1},\sigma);\nu\right)}\,du\\
		\nonumber	&=&\sigma^{-m_{i}}\bm\psi_{m_{i}}\left(\bm W_{\bm y_{i}}^{\nu}(\bm\beta_{1},\sigma);\nu,\bm R(\rho_{y})_{(\bm y_{i})}\right)\prod_{j=1}^{m_{i}}\frac{\phi\left(\frac{ y_{ij}-\bm x_{ij1}^{'}\bm\beta_{1}}{\sigma}\right)}{\psi\left(W_{y_{ij}}^{\nu}(\bm\beta_{1},\sigma);\nu\right)}\\
		\nonumber&\times&\int_{t_{i}}^{\infty}\psi_{1}\left(W_{u}^{\nu}(\bm\beta_{2},r);\nu_{i}^{t_{i}|\bm y_{i}},\mu_{i}^{t_{i}|\bm y_{i}},\left(\sigma_{i}^{t_{i}|\bm y_{i}}\right)^2 \right)\frac{f_{T_{i}^{*}}(u;\bm x_{i2}^{'}\bm\beta_{2},r)}{\psi\left(W_{u}^{\nu}(\bm\beta_{2},r);\nu\right)}\,du\\
		\nonumber	&=&\sigma^{-m_{i}}\bm\psi_{m_{i}}\left(\bm W_{\bm y_{i}}^{\nu}(\bm\beta_{1},\sigma);\nu,\bm R(\rho_{y})_{(\bm y_{i})}\right)\prod_{j=1}^{m_{i}}\frac{\phi\left(\frac{ y_{ij}-\bm x_{ij1}^{'}\bm\beta_{1}}{\sigma}\right)}{\psi\left(W_{y_{ij}}^{\nu}(\bm\beta_{1},\sigma);\nu\right)}\\
			\nonumber	&\times&\int_{W_{t_{i}}^{\nu}(\bm\beta_{2},r)}^{\infty}\psi_{1}\left(w;\nu_{i}^{t_{i}|\bm y_{i}},\mu_{i}^{t_{i}|\bm y_{i}},\left(\sigma_{i}^{t_{i}|\bm y_{i}}\right)^2\right)\,dw	
	\end{eqnarray*}
	\begin{eqnarray}
		\nonumber &=&\sigma^{-m_{i}}\bm\psi_{m_{i}}\left(\bm W_{\bm y_{i}}^{\nu}(\bm\beta_{1},\sigma);\nu,\bm R(\rho_{y})_{(\bm y_{i})}\right)\prod_{j=1}^{m_{i}}\frac{\phi\left(\frac{ y_{ij}-\bm x_{ij1}^{'}\bm\beta_{1}}{\sigma}\right)}{\psi\left(W_{y_{ij}}^{\nu}(\bm\beta_{1},\sigma);\nu\right)}\\
		&\times&\Psi\left(-\frac{W_{t_{i}}^{\nu}(\bm\beta_{2},r)-\mu_{i}^{t_{i}|\bm y_{i}}}{\sigma_{i}^{t_{i}|\bm y_{i}}};\nu_{i}^{t_{i}|\bm y_{i}}\right),\label{tlik2}
	\end{eqnarray}
	
	where \[\nu_{i}^{t_{i}|\bm y_{i}}=\nu+m_{i},\]
	\[\mu_{i}^{t_{i}|\bm y_{i}}=\bm R(\rho_{ty})_{(t_{i})(\bm y_{i})}\bm R(\rho_{y})_{(\bm y_{i})}^{-1}\bm W_{y_{i}}^{\nu}(\bm\beta_{1},\sigma)\] and
	\[\left(\sigma_{i}^{t_{i}|\bm y_{i}}\right)^{2}=\frac{\left(\nu+\bm W_{y_{i}}^{\nu}(\bm\beta_{1},\sigma)^{'}\left(\bm R(\rho_{y})_{(\bm y_{i})}\right)^{-1}\bm W_{y_{i}}^{\nu}(\bm\beta_{1},\sigma)\right)\left(1-\bm R(\rho_{ty})_{(t_{i})(\bm y_{i})}\bm R(\rho_{y})_{(\bm y_{i})}^{-1}\bm R(\rho_{ty})_{(\bm y_{i})(t_{i})}\right)}{\nu_{i}^{t_{i}|\bm y_{i}}}.\] We use the fact that the marginal and conditional distributions are closed under the multivariate $t$ distribution (Ding, 2016\cite{din16}).

\subsection{Log-likelihood and maximisation}

The log-likelihood function under both the multivariate Gaussian and $t$ copula,  plugging in expression (\ref{Gaulik1}) and (\ref{Gaulik2}) or (\ref{tlik1}) and (\ref{tlik2}), can be written analytically as:
\begin{eqnarray}
	\displaystyle
	l(\theta)&=&\sum_{i}^{n}\delta_{i}\mbox{log}f_{T_{i}^{*},\bm y_{i}}(t_{i},\bm y_{i})+\sum_{i}^{n}(1-\delta_{i})\mbox{log}f_{T_{i}^{*},\bm y_{i}}(T_{i}^{*}>t_{i},\bm y_{i}).
	\label{3}
\end{eqnarray}
This formulation applies to all subjects, unlike the approach in Ganjali and Baghfalaki (2015)\cite{gan15} requiring four different categories of log-likelihood functions. The score function of (\ref{3}) can be solved numerically and we apply a Newton-type algorithm (Dennis, \textit{et al,} 1983\cite{den83}) and the approach of Nelder and Mead (1965)\cite{nel65} for maximising (\ref{3}), which are implemented by the \verb|nlm| and \verb|optim| functions in \verb|R|. Under the multivariate Gaussian copula, the two sub-models are equivalent to being modelled independently if  $\bm R(\rho_{ty})_{(t_{i})(\bm y_{i})}$ in correlation matrix (\ref{RG}) is a vector of zeros, but this is not the case under the multivariate $t$ copula. Analytical solutions for estimating some of the parameters are available when the true event times of all subjects are observed under the multivariate Gaussian copula. Details are provided in the Appendix A. The multivariate $t$ copula, however, does not have this nice property. In later Sections, we use $\bm R=\bm R(\cdot)$ for simplicity.

\section{Simulation studies}
 Unlike the observable marginal variables, the correlation structure between the marginals is unobservable and is more likely to be misspecified.  We would like to investigate the consequences of misspecifying the copula when the marginal distributions are correctly specified. In the two simulation studies, $N=500$ Monte Carlo samples with sample size $n=200$ are generated under  the same marginals, but with the multivariate Gaussian copula and $t$ copula with 4 degrees of freedom,respectively for the joint distributions.
	A longitudinal process with five measurements at $t_{1}=0,$ $t_{2}=2,$ $t_{3}=6,$ $t_{4}=12$ and $t_{5}=18$ is generated from the following model:
	\begin{eqnarray*}
		\displaystyle
		y_{ij}=\beta_{01}+\beta_{11}t_{j}+\beta_{21}t_{j}x_{i1}+\beta_{31}x_{i2}+\beta_{41}x_{i3}+\beta_{51}x_{i4}+\varepsilon_{ij},\mbox{  }\mbox{   }\mbox{   }i=1,...,n,\mbox{  }j=1,...,5,
	\end{eqnarray*}
	where $x_{i1},$ $x_{i2},$ $x_{i3}$ and $x_{i4}$ follow Bernoulli distributions with probabilities $(0.5,0.5,0.31,0.21)$ taking value 1 and $\varepsilon_{ij}\sim N(0,\sigma^{2})$.  An associated event time process with a Weibull proportional hazard model is specified as:
	\begin{eqnarray*}
		\displaystyle 
		h_{i}(t)=rt^{r-1}\mbox{exp}\left(\beta_{02}+\beta_{12}x_{i1}+\beta_{22}x_{i2}+\beta_{32}x_{i3}+\beta_{42}x_{i4}\right).
	\end{eqnarray*}
	
	Subjects who do not have all the scheduled measurements are counted as dropout and those whose true event times are not observed are counted as censored.
	Independent exponential censoring processes with different rates of 0.000, 0.0225 and 0.0610 are introduced to control the dropout (censoring) rate to be roughly  40\% (60\%), 60\% (67\%) and 80\% (75\%).  The outputs of parameter estimation based on the 500 Monte Carlo samples with 40\% (60\%) dropout (censoring) rate are listed in Tables \ref{simGau1} and \ref{simt1}, where SE and SD represent the model-based standard errors from the inverse Hessian matrix and the empirical standard deviations, respectively, while CP stands for the coverage probabilities from the model-based 95\% normal confidence intervals. The root mean square error of parameter $\theta$ is defined as $RMSE(\theta)=\sqrt{\frac{1}{N}\sum_{j=1}^{N}\left(\hat{\theta}_{j}-\theta\right)^{2}}$ with $\hat{\theta}_{j}$ being the parameter estimate of $\theta$ for the $j$th sample. The outputs for the other dropout rates are summarised in Appendix B.
	
	\subsection{Simulation study 1}
	The multivariate Gaussian copula joint model with an exchangeable correlation structure within the longitudinal process with $\rho_{y}=0.4$ and a constant correlation between the two sub-models with $\rho_{ty}=0.6,$ is used to generate the data. In this case, the longitudinal process follows a multivariate normal distribution. We perform the estimation process under the multivariate Gaussian copula joint model and the multivariate $t$ copula joint model with 3, 4, 40 and 100 degrees of freedom when the marginal distributions and the correlation matrix $\bm R_{(t,\bm y)}$ are correctly specified.
	
	According to Table \ref{simGau1}, when the multivariate Gaussian copula joint model is correctly specified, the outputs of parameter estimations are generally accurate with low biases. The RMSEs are almost identical to or only slight larger than the SDs, which again indicates very low biases. The model-based standard errors from the inverse Hessian matrix (SE) and the empirical standard deviations (SD) are very similar and the coverage probabilities are all around 95\%. When the copula function is misspecified as the multivariate $t$ distribution with 3 degrees of  freedom $(df=3)$, obvious biases can be observed in the estimates of parameters for the survival process. The SEs are generally underestimated and smaller than the SDs except for the correlation parameters $\rho_{ty}$ and $\rho_{y}.$ The coverage probabilities for the parameters of the survival process are obviously lower than the 95\% 
	\begin{landscape}
		\vspace*{\fill}
		\begingroup
		\setlength{\tabcolsep}{6pt} 
		\renewcommand{\arraystretch}{1.18} 
		\begin{table}[H]
				\tbl{\textbf{{\scriptsize Data are generated by the multivariate Gaussian copula  joint model with 40\% dropout (60\% censoring) rate and parameters are estimated by the multivariate Gaussian copula  joint model and multivariate $t$ copula  joint model with 3, 4, 40 and 100 degrees of freedom when the marginal distributions and $\bm R_{(t,\bm y)}$ are correctly specified.}}}
				{\begin{tabular}{lccccccccccccccccc}
						\toprule
						True value &$\beta_{01}$&$\beta_{11}$&$\beta_{21}$&$\beta_{31}$&$\beta_{41}$&$\beta_{51}$&$\beta_{12}$&$\beta_{22}$&$\beta_{32}$&$\beta_{42}$&$\beta_{52}$&$r$&$\sigma$&$\rho_{ty}$  &$\rho_{y}$
						\\
						& 5         & 1         & 2                 & 1             &  -2         &   -1        & -5        &-4      & -2       & 2   &1    & 2       &  3       &  0.6         & 0.4      
						\\
						\hline
						Gaussian copula &                   &                   &                    &                   &                   &                   &                    &                    &                     &                   &
						\\
						\cdashline{2-16}
						Est.        &5.013 & 0.999  & 2.002  & 0.986 &-2.009 & -1.004 & -5.118 & -4.144 &-2.046 &2.048 &1.037 & 2.047 & 2.976  & 0.601  & 0.391
						\\
						SE         & 0.280 & 0.024 &  0.027 &  0.307 & 0.352 & 0.396  & 0.458  & 0.397  & 0.237 & 0.254 &0.267 & 0.169 & 0.092  & 0.036 & 0.038 
						\\
						SD         & 0.267  & 0.024 & 0.028  & 0.314  &0.334  &0.393   &0.467    &0.412  &0.251  &0.254  &0.272 &0.173  &0.091   & 0.036  &0.038  
						\\
						RMSE    & 0.267 & 0.024  & 0.028  &  0.314 &0.334  &0.393   &0.481    & 0.436 & 0.255 & 0.259 &0.275 &0.179  & 0.094 & 0.036  & 0.039
						\\
						CP	      & 0.954 & 0.946 & 0.938  & 0.944  & 0.960  & 0.948  &0.962    &0.942  & 0.942 &0.948 &0.958 & 0.960 & 0.924 &0.948   & 0.942   
						\\
						\cdashline{1-16}
						$t$ copula ($df=3$)  & &  &   &   &    &   &  &  &  &  &
						\\
						\cdashline{2-16}
						Est.     &5.041   & 0.997 &2.002  &0.993   &-2.004 &-0.995 &-4.604 &-3.666 &-1.834 & 1.820  &0.915 & 1.830  & 3.404  &0.579 &0.364
						\\
						SE       &0.282   & 0.024 &  0.027&  0.310 & 0.356 & 0.398 & 0.393  & 0.324  & 0.209 & 0.223 &0.239 &0.145   & 0.109  & 0.043 &0.045
						\\
						SD       & 0.296  & 0.027 & 0.031 &  0.350 & 0.371 & 0.435 & 0.428  & 0.380  & 0.239  & 0.236 &0.260 & 0.157  & 0.116  & 0.040 & 0.040 
						\\
						RMSE  & 0.298  & 0.027 &0.031  & 0.350  & 0.371 & 0.435  &0.583  & 0.506  & 0.291  & 0.296  &0.274  & 0.231 & 0.420 & 0.045 & 0.054
						\\
						CP	    & 0.940  & 0.924 & 0.900 & 0.914  & 0.942 & 0.930 &0.742  &0.728   &0.810    & 0.852  &0.924  & 0.720 & 0.030 & 0.958 & 0.896
						\\
						\cdashline{1-16}
						$t$ copula ($df=4$)  & &  &   &   &    &   &  &  &  &  &
						\\
						\cdashline{2-16}
						Est.     &5.030   & 0.998 &2.002  &0.993  &-2.005 &-0.996 &-4.786 &-3.834 &-1.909 & 1.900 &0.955 & 1.908  & 3.254  &0.583 &0.369
						\\
						SE       &0.281   & 0.024 &  0.027&  0.308 & 0.354 & 0.396 & 0.414  & 0.346  & 0.218 & 0.233 &0.248 &0.153   & 0.105  & 0.042 &0.044
						\\
						SD      & 0.286  & 0.026 & 0.030 &  0.339 & 0.359 & 0.419 & 0.444  & 0.394 & 0.245 & 0.242 &0.265  & 0.164  & 0.106 & 0.039 & 0.039 
						\\
						RMSE & 0.287  & 0.026  &0.030  & 0.339 & 0.358  & 0.419 &0.493  & 0.428 & 0.261  & 0.262 &0.269  & 0.188  & 0.276 & 0.042 & 0.050
						\\
						CP	    & 0.946 & 0.934  & 0.924 & 0.926 & 0.940  & 0.942 &0.864  &0.856 &0.888   & 0.906 &0.942  & 0.856 & 0.310  & 0.956 & 0.916
						\\
						\cdashline{1-16}
						$t$ copula ($df=40$)   & &  &   &   &    &   &  &  &  &  &
						\\
						\cdashline{2-16}
						Est.    & 5.011   &0.999  & 2.002   &0.989  & -2.007 & -1.001 &-5.113 & -4.138 & -2.046 &2.043 &1.035  &2.046   &2.982   &0.600  &0.390
						\\
						SE      & 0.280  & 0.024 & 0.027  &  0.307 &  0.352 & 0.396  & 0.457 & 0.394  & 0.237  & 0.253 &0.266  & 0.169   & 0.093 & 0.037 &0.039 
						\\
						SD      & 0.267  & 0.024 &  0.028 &  0.314 & 0.334  &  0.393 & 0.463 & 0.414  & 0.251   & 0.251 & 0.271  & 0.171   & 0.091  & 0.036 & 0.038
						\\
						RMSE & 0.267  & 0.024 & 0.028  & 0.314   & 0.334  & 0.393  & 0.476 & 0.436  & 0.255 & 0.255  &0.273  & 0.177  & 0.093   & 0.036 & 0.040
						\\
						CP	    & 0.956 & 0.938 & 0.934  & 0.944  & 0.962  & 0.946   &0.954  &0.944  & 0.942  & 0.952 &0.960  & 0.960  & 0.938  & 0.948  & 0.944  
						\\
						\cdashline{1-16}
						$t$ copula ($df=100$)   & &  &   &   &    &   &  &  &  &  &
						\\
						\cdashline{2-16}
						Est.    &5.010    &0.999  & 2.002  &0.988  & -2.008 & -1.001  &-5.117 & -4.141 & -2.047&2.048   &1.035  &2.047  &2.977   &0.601    &0.391
						\\
						SE      & 0.280  & 0.024 &  0.027 &  0.307 &  0.352 & 0.396   & 0.457 & 0.395 & 0.237  & 0.254  &0.267 & 0.169 & 0.092  & 0.036   & 0.039
						\\
						SD      & 0.267  & 0.024 &  0.028 &  0.315 & 0.334  &  0.392  & 0.465 & 0.414 & 0.251  & 0.254   &0.271 & 0.172 & 0.090   &  0.036 & 0.038 
						\\
						RMSE & 0.267  & 0.024 & 0.028  &0.315   &0.334   & 0.392   & 0.479 & 0.437 & 0.255  & 0.259  & 0.274 & 0.179 &0.093   &  0.036  & 0.039  
						\\
						CP	    & 0.956 & 0.944 & 0.934  & 0.944 & 0.960  & 0.944   &0.960  &0.942  &0.942   & 0.952  &0.958  & 0.958 & 0.934 & 0.946   & 0.940
						\\
						\hline
				\end{tabular}}
				\label{simGau1}
		\end{table}
		\endgroup
		\vspace*{\fill}
	\end{landscape}

	\begin{landscape}
		\vspace*{\fill}
		\begingroup
		\setlength{\tabcolsep}{6pt} 
		\renewcommand{\arraystretch}{1.18} 
		\begin{table}[H]
				\tbl{\textbf{{\scriptsize Data are generated by the multivariate $t$ copula  joint model with 4 degrees of freedom and 40\% dropout (60\% censoring) rate and parameters are estimated by the multivariate Gaussian copula  joint model and multivariate $t$ copula  joint model with 3, 4 and 5 degrees of freedom when the marginal distributions and $\bm R_{(t,\bm y)}$ are correctly specified.}}}
				{\begin{tabular}{lccccccccccccccccc}
						\toprule
						True value &$\beta_{01}$&$\beta_{11}$&$\beta_{21}$&$\beta_{31}$&$\beta_{41}$&$\beta_{51}$&$\beta_{12}$&$\beta_{22}$&$\beta_{32}$&$\beta_{42}$&$\beta_{52}$&$r$&$\sigma$&$\rho_{ty}$  &$\rho_{y}$
						\\
						& 5         & 1         & 2                 & 1             &  -2         &   -1        & -5        &-4      & -2       & 2   &1    & 2       &  3       &  0.6         & 0.4      
						\\
						\hline
						$t$ copula ($df=4$) &                   &                   &                    &                   &                   &                   &                    &                    &                     &                   &
						\\
						\cdashline{2-16}
						Est.       &5.006  & 1.000  & 2.000  & 0.992 &-2.000 & -0.996 & -5.118 & -4.111 &-2.047 &2.055  &1.041 & 2.046 & 2.982  & 0.603  & 0.394
						\\
						SE         & 0.256 & 0.021 &  0.024 &  0.280 & 0.321  & 0.360  & 0.422  & 0.352  & 0.215 & 0.231 &0.241 & 0.156 & 0.097  & 0.040 & 0.042
						\\
						SD         & 0.247 & 0.022 & 0.025  & 0.276  &0.322  &0.360   &0.425    &0.358  &0.208  &0.239 &0.245 &0.156  &0.090   & 0.040  &0.041  
						\\
						RMSE    & 0.247 & 0.022 & 0.025  &  0.276 &0.322  &0.360   &0.441    & 0.374  & 0.213  & 0.245&0.248 &0.163  & 0.092 & 0.040  & 0.041
						\\
						CP	      & 0.954 & 0.954 & 0.936  & 0.958  & 0.950 & 0.952  &0.946   &0.954   & 0.966 &0.950  &0.942 & 0.946 & 0.954 &0.948   & 0.956   
						\\
						\cdashline{1-16}
						Gaussian copula  & &  &   &   &    &   &  &  &  &  &
						\\
						\cdashline{2-16}
						Est.     &5.012   & 0.999 &2.005  &0.992   &-1.995  &-0.996 &-5.130 &-4.101  &-2.042 & 2.060 & 1.048 & 2.047  & 2.973 &0.593 &0.386
						\\
						SE       &0.280  & 0.024 &  0.027 &  0.306 & 0.352 & 0.394 & 0.462  & 0.391  & 0.239 & 0.258 &0.268  &0.170   & 0.091  & 0.037 &0.038
						\\
						SD       & 0.275  & 0.026 & 0.029 & 0.296 & 0.354 & 0.407 & 0.476  & 0.396  & 0.235 & 0.269  &0.285  & 0.172  & 0.107  & 0.047 & 0.046 
						\\
						RMSE  & 0.275  & 0.026 &0.029  & 0.296 & 0.353  & 0.406 &0.493  & 0.409 & 0.238  & 0.276  &0.289  & 0.178  & 0.110  & 0.048 & 0.048
						\\
						CP	    & 0.948  & 0.940 & 0.946 & 0.956 & 0.944  & 0.934 &0.946  &0.954 &0.962   & 0.940 &0.922  & 0.946 & 0.862  & 0.882 & 0.876
						\\
						\cdashline{1-16}
						$t$ copula ($df=5$)   & &  &   &   &    &   &  &  &  &  &
						\\
						\cdashline{2-16}
						Est.    & 5.005  &1.000   & 2.000  &0.991   & -2.003 & -0.997 &-5.186 & -4.171 & -2.072 &2.085  &1.057  &2.073   &2.935    &0.605  &0.397
						\\
						SE      & 0.260  & 0.022 & 0.024  &  0.284 &  0.326 & 0.365  & 0.434 & 0.365  & 0.221  & 0.238 &0.247  & 0.160   & 0.096  & 0.039 &0.042
						\\
						SD      & 0.248  & 0.022 &  0.025 &  0.273 & 0.323  &  0.362 & 0.434 & 0.365  & 0.208  & 0.243 & 0.249  & 0.159  & 0.090  & 0.040 & 0.041
						\\
						RMSE & 0.248  & 0.022 & 0.025  & 0.273   & 0.323  & 0.362  & 0.472 & 0.403  & 0.220 & 0.257  &0.255   & 0.175  & 0.110   & 0.040 & 0.041
						\\
						CP	    & 0.960 & 0.952 & 0.942  & 0.962  & 0.952  & 0.954   &0.946  &0.954  & 0.968  & 0.950 &0.938  & 0.936  & 0.894  & 0.938  & 0.956  
						\\
						\cdashline{1-16}
						$t$ copula ($df=3$)  & &  &   &   &    &   &  &  &  &  &
						\\
						\cdashline{2-16}
						Est.    &5.017   &1.000  & 1.999   &0.992  & -2.006 & -0.996  &-4.981 & -3.984 & -1.988&1.997   &1.011   &1.987   &3.079   &0.599   &0.389
						\\
						SE      & 0.251  & 0.021  &  0.023 &  0.274 &  0.315 & 0.353    & 0.401 & 0.331  & 0.206  & 0.221  &0.231  & 0.148  & 0.099  & 0.041  & 0.043
						\\
						SD      & 0.249  & 0.022 &  0.024 &  0.278 & 0.326  &  0.360   & 0.411 & 0.343 & 0.200  & 0.233  &0.237  & 0.151  & 0.092  &  0.040 & 0.040 
						\\
						RMSE & 0.249  & 0.022 & 0.024  &0.278   &0.325   & 0.360    & 0.411  & 0.343 & 0.200  & 0.233 & 0.237  & 0.151  &0.121    &  0.040 & 0.042  
						\\
						CP	    & 0.948 & 0.944 & 0.934  & 0.948  & 0.954  & 0.948    &0.944  &0.936  &0.958   & 0.936  &0.940  & 0.942 & 0.906  & 0.956  & 0.962
						\\
						\hline
				\end{tabular}}
				\label{simt1}
		\end{table}
		\endgroup
		\vspace*{\fill}
	\end{landscape}

	\noindent nominal level. For parameter $\sigma,$ the  coverage probability is only 3\%. The effect of misspecification is less significant if the degrees of freedom of the multivariate $t$ copula increases to 4. The outputs under $df=40$ are  very close to the outputs under the multivariate Gaussian copula joint model.  If we further increase $df$ from 40 to 100, the outputs are almost identical to the results under the multivariate Gaussian copula joint model, although the improvement is much less obvious as increasing $df$ from 3 to 4. As the number of parameters are the same under these five models, the log-likelihood can be compared directly to assess the goodness of fit. The multivariate Gaussian copula joint model outperforms the multivariate $t$ copula joint model with $df=3,$ 4, 40 and 100 in 500, 500, 372 and 314 out of 500 Monte Carlo samples, which means the log-likelihood values are sufficient to select the true model out of the five candidates.

	\subsection{Simulation study 2}
	
	The multivariate $t$ copula joint model with 4 degrees of freedom with the same marginal distributions and correlation matrix $\bm R_{(t,\bm y)}$ as in simulation study 1, are used to generate the data in simulation study 2. In this case, although a single longitudinal measurement follows the normal distribution, the longitudinal process as a whole for a subject is not jointly multivariate normally distributed. We perform the estimation process under the multivariate Gaussian copula joint model and the multivariate $t$ copula joint model with 3, 4 and 5 degrees of freedom when the marginal distributions and the correlation matrix $\bm R_{(t,\bm y)}$ are correctly specified.
	
	According to Table \ref{simt1}, when we correctly specify the copula as multivariate $t$ copula with $df=4,$ the parameters are estimated with very low biases, similar values of SEs, SDs and RMSEs. The coverage probabilities are very close to the 95\% nominal level. The parameter estimation under misspecification is rather robust, especially for the parameters of the survival process and the regression parameters for the longitudinal process, since values of SEs, SDs and RMSEs are close, and the value of the CPs is around 95\%. However, the estimation under the multivariate Gaussian copula joint model may be less efficient as we observe generally larger SEs and SEs.  The multivariate $t$ copula joint model with $df=4$ outperforms the multivariate Gaussian copula joint model and the multivariate $t$ joint model with $df=3$ and 5 in 500, 375, 380 out of 500 Monte Carlo samples, respectively. As in simulation study 1, comparing the the log-likelihood values is sufficient to select the true model out of the four candidates. 
	
	According to Tables \ref{simGau2} to \ref{simt3} in Appendix B, if we increase the dropout (censoring) rate, the values of SEs and SDs increase as there are fewer longitudinal measurements, which means less information and the impact of misspecification of the copula function becomes less obvious. Indeed, if there are fewer or, in the most extreme case, no longitudinal measurements, joint modelling is less useful or even not necessary and selection of the copula function becomes less important. This conclusion can be drawn for both simulation studies 1 and 2. In addition, we also considered other correlation structures such as $\bm R_{(\bm y)}$ to be an AR1 and $\bm R_{(t)(\bm y)}=\rho_{ty}^{(6-j)},$ $j=1,..,5,$ with $\rho_{y}=0.5$ and $\rho_{ty}=0.75$ in simulation studies 1 and 2 and we can conclude similar conclusions from the outputs (the results are not presented here due to space limitations).
	
	Although the log-likelihood expressions (\ref{Gaulik1}), (\ref{Gaulik2}), (\ref{tlik1}) and (\ref{tlik2}) are derived under normal marginals for longitudinal process (the normality assumption is the most common assumption for continuous biomarker measurements), they can be easily adapted to other marginals distributions. For example, we also simulated data by the multivariate $t$ copula with $df=4$ under $t$ marginals with $df=4$ for the longitudinal process and a Weibull marginal for the event time and the estimation results were good if the model was correctly specified. Due to space limitations, we do not include these results.

\subsection{Dynamic prediction for survival time}
The prediction of the survival probability for subjects with some baseline covariates and longitudinal measurements is also of interest after a joint model is fitted. Even though there are no subject specific random effects to distinguish the individual trajectories from the population trend, this copula approach for joint modelling is capable of predicting survival probabilities at an individual level.

Suppose a copula joint model is fitted based on a random sample of $n$ subjects $\mathcal{\bm D}_{n}=\left\{T_{i},\delta_{i},\bm y_{i};i=1,...,n\right\}.$ Predictions of survival probabilities at time $u>t$ for a new subject $i$ which has a set of longitudinal measurements $\mathcal{\bm Y}_{i}(t)=\left\{y_{i}(s);0\leq s<t\right\}$ up to $t$ and a vector of baseline covariates $\bm x_{i2}$ are given by:
\begin{eqnarray}
	\nonumber	\pi_{i}(u|t)&=&P(T_{i}>u|T_{i}>t,\mathcal{\bm Y}_{i}(t),\bm x_{i2},\mathcal{\bm D}_{n};\bm\theta)=P(T_{i}>u|T_{i}>t,\mathcal{\bm Y}_{i}(t),\bm x_{i2};\bm\theta)\\ 
	&=&\frac{P(T_{i}>u|\mathcal{\bm Y}_{i}(t),\bm x_{i2},\bm b_{i};\bm\theta)}{P(T_{i}>t|\mathcal{\bm Y}_{i}(t),\bm x_{i2},\bm b_{i};\bm\theta)}.\label{predt}
\end{eqnarray}

	For the multivariate Gaussian copula, (\ref{predt}) becomes
	\begin{eqnarray}
		\Phi\left(\frac{\Phi^{-1}\left(S_{T_{i}}(u;\bm x_{i2}^{'}\bm\beta_{2},r)\right)+\mu_{i}^{u|\bm y_{i}}}{\sigma_{i}^{u|\bm y_{i}}}\right)\bigg/\Phi\left(\frac{\Phi^{-1}\left(S_{T_{i}}(t;\bm x_{i2}^{'}\bm\beta_{2},r)\right)+\mu_{i}^{t|\bm y_{i}}}{\sigma_{i}^{t|\bm y_{i}}}\right),\label{Gcopredt}
	\end{eqnarray}
	which reduces to $\displaystyle\frac{S_{T_{i}}(u;\bm x_{i2}^{'}\bm\beta_{2},r)}{S_{T_{i}}(t;\bm x_{i2}^{'}\bm\beta_{2},r)}=\displaystyle\mbox{exp}\left\{-\int_{t}^{u}h_{i}(s)ds\right\}$ given $\bm R_{(t)(\bm y_{i})}=\bm0$  and $\bm R_{(u)(\bm y_{i})}=\bm0.$ This means the joint modelling reduces to separately fitting of the two sub-models if the correlation between them is zero under the multivariate Gaussian copula joint model. The definitions of conditional mean and standard deviation are the same as in (\ref{Gaulik2}). 
	
	For the multivariate $t$ copula, (\ref{predt}) becomes
	\begin{eqnarray}
		\Psi\left(\frac{\Psi^{-1}\left(S_{T_{i}}(u;\bm x_{i2}^{'}\bm\beta_{2},r);\nu\right)+\mu_{i}^{u|\bm y_{i}}}{\sigma_{i}^{u|\bm y_{i}}};\nu^{u|\bm y_{i}}\right)\bigg/\Psi\left(\frac{\Psi^{-1}\left(S_{T_{i}}(t;\bm x_{i2}^{'}\bm\beta_{2},r);\nu\right)+\mu_{i}^{t|\bm y_{i}}}{\sigma_{i}^{t|\bm y_{i}}};\nu^{t|\bm y_{i}}\right),\label{tcopredt}
	\end{eqnarray}
	where the definition of conditional mean, standard deviation and degrees of freedom are the same as in (\ref{tlik2}). 
	
	To assess the prediction accuracy of the proposed model, we calculate the mean residual lifetime $\displaystyle \bar{r}_{i}^{t}=E\left[T_{i}-t| T_{i}>t\right]=\int_{t}^{\infty}\pi_{i}(u|t)\,du,$ $i\in \{\mbox{subjects at risk at time } t\}$, at $t=$ 2, 6, 12 and 18 for 1000 new subjects generated as in simulation studies 1 and 2, respectively. The parameter estimates $\hat{\bm\theta}$ from Tables \ref{simGau1} and \ref{simt1} are substituted into (\ref{Gcopredt}) and (\ref{tcopredt}) for calculation. We also calculate the mean residual time by fitting the correct survival sub-model alone, which corresponds to $\bm R_{(t)(\bm y_{i})}=\bm0$  and $\bm R_{(u)(\bm y_{i})}=\bm0$ under the multivariate Gaussian copula joint model. The model with mean residual lifetimes closer to the true residual lifetimes $r_{i}^{t}=t_{i}-t$ ($t_{i}$ is the true event time) is better in terms of predicting survival time. 
	
	When joint data is generated under the multivariate Gaussian copula joint model as in simulation study 1, the average absolute differences between the mean and true residual lifetimes, measured by $\displaystyle\sum_{i}|\bar{r}_{i}^{t}-r_{i}^{t}|/\{\# \mbox{ of subjects at risk at time } t\},$ for the correctly specified model, the multivariate $t$ copula joint model with $df=4$ and survival sub-model alone are 17.934, 18.158 and 26.554 at $t=2;$ 18.301, 20.228 and 29.446 at $t=6;$ 20.150, 20.244 and 33.528 at $t=12;$ 20.945, 21.148 and 37.017 at $t=18.$ The correctly specified joint model provides the most accurate predictions on average but the performance of the multivariate $t$ copula joint model with $df=4$ is very close and they are both  much better than the survival sub-model alone. As $t$ increases, the discrepancy between mean and true residual lifetimes is increasing, especially for when the survival sub-model is used alone, and this is because the survival times are very right skewed. For example, at $t=18,$ the subjects at risk tend to have longer survival times (e.g. more than 100), which makes predicting survival time more difficult and less accurate. It is also noticeable that the differences in predicted survival time between the survival sub-model and the other two copula joint models are increasing when $t$  increases, because the copula joint models can utilise more information from the longitudinal process for making prediction at later timepoints, but the survival sub-model alone cannot. The same results can be drawn when joint data is generated under the multivariate $t$ copula joint model with $df=4$ as in the simulation study 2.
	
	Table \ref{comsur} summarises the percentages of better predicting models  between the correctly specified joint models, misspecified joint models and correctly specified survival sub-model alone for these 1000 subjects in the two simulation studies. At each time point, there are more correctly specified joint models outperforming the misspecified joint models and survival sub-model. However, the misspecified joint models still offer a good improvement in accuracy for predicting survival time compared with the survival sub-model, which suggests it is always preferable to use the joint model than the survival sub-model alone for predicting survival probabilities, even if we do not know the true correlation between the sub-models.
	
	We also notice that the survival sub-model still provides better prediction for a certain number of subjects (about 1/3) in survival time. Some subjects have longitudinal trajectories very close to the population trend, in this case the correctly specified joint model, misspecified joint model and survival sub-model provide very similar prediction, as presented by subject 162 from the AIDS data in Figure \ref{cd4bigplot}, thus the survival sub-model could provide marginally better predictions than the other candidates.
	\begingroup
	\setlength{\tabcolsep}{6pt} 
	\renewcommand{\arraystretch}{1.18} 
	\begin{table}[H]
			\tbl{\textbf{{\scriptsize Percentages for the fitted multivariate $t$ copula joint model with $df=4,$  multivariate Gaussian copula  joint model and  survival sub-model alone indicating better accuracy of predicting survival probabilities for 1000 new subjects generated as in simulation studies 1 and 2, respectively.}}}
			{\begin{tabular}{lcccc}
					\toprule
					&$t=2$&$t=6$&$t=12$& $t=18$      
					\\
					\midrule
					\multirow{2}{*}{Data generated by the multivariate}&&&&  
					\\  
					\\    
					Gaussian copula  joint model&&&&  
					\\
					\cdashline{2-5}
					\\
					Gaussian copula vs. $t$ copula $(df=4)$       &  54.39\% vs. 45.61\%   &  58.36\% vs. 42.64\% & 52.12\% vs. 47.88\% & 55.46\% vs. 44.54\% 
					\\            
					Gaussian copula vs. Survival sub-model          & 66.36\% vs. 33.64\%   & 66.43\% vs. 33.57\%   & 67.00\% vs. 33.00\% & 68.84\% vs. 31.16\%
					\\            
					$t$ copula $(df=4)$  vs. Survival sub-model   & 66.05\% vs. 33.95\%   & 65.96\% vs. 34.04\%   & 66.43\% vs. 33.57\% & 67.86\% vs. 32.14\%
					\\
					\\
					\cdashline{2-5}                                                            
					Number of  subjects at risk time $t$                   &969                &855                &706               &613
					\\
					\midrule
					\multirow{2}{*}{Data generated by the multivariate}&&&&
					\\
					\\  
					$t$ copula joint model with $df=4$ &&&&     
					\\     
					\cdashline{2-5}
					\\
					$t$ copula $(df=4)$ vs. Gaussian copula      &  54.30\% vs. 45.70\% &  53.62\% vs. 46.38\%  & 55.69\% vs. 44.31\%  & 60.10\% vs. 39.90\% 
					\\            
					$t$ copula $(df=4)$ vs. Survival sub-model    & 67.93\% vs. 32.07\%  & 70.03\% vs. 29.97\%    & 72.22\% vs. 27.78\% & 72.31\% vs. 27.69\%
					\\            
					Gaussian copula vs. Survival sub-model          & 66.70\% vs. 33.30\%  & 68.54\% vs. 31.46\%   & 70.00\% vs. 30.00\% & 71.01\% vs. 28.99\%
					\\
					\\
					\cdashline{2-5}
					Number of  subjects at risk time $t$                 &976                &871                &720               &614
					\\
					\bottomrule
			\end{tabular}}
			\label{comsur}
	\end{table}
	\endgroup

\section{Application to the AIDS data}
The AIDS data (Goldman \textit{et al.,} 1996\cite{gol96}) is included in many \verb|R| packages such as \verb|JM| (Rizopoulos, 2010\cite{riz10}) or \verb|joineR| (Philipson, {\it et al,} 2017\cite{phi17}). The dataset comprises the square root of CD4 cell counts per cubic millimetre ($mm^{3}$
) in blood for 467 subjects with advanced human immunodeficiency virus infection at study entry and later on at 2, 6, 12, and 18 months. The CD4 cell count is an important indicator for the progression from HIV infection to acquired immune
deficiency syndrome (AIDS).  A higher CD4 level indicates a stronger immune system, thus less likely to lead to AIDS diagnosis or death. Some other covariates such as gender, randomly assigned treatment by didanosine (ddI) or zalcitabine (ddC) are also recorded at baseline. The main purpose of this study is to compare the efficacy and safety of ddI and ddC. 
By the end of the study 188 patients had died, resulting in about 59.7\%
censoring, and out of the 2335 planned measurements, 1405 were actually recorded, leading to 39.8\%  missing responses.

The longitudinal process is specified as:
\begin{eqnarray}
	\displaystyle
	y_{ij}=\beta_{01}+\beta_{11}t_{j}+\beta_{21}t_{j}drug_{i}+\beta_{31}gender_{i}+\beta_{41}prevOI_{i}+\beta_{51}AZT_{i}+\varepsilon_{ij},\\
	\nonumber i=1,...,467,\mbox{  }1\leq j\leq5,
	\label{aidslong}
\end{eqnarray}
where $\varepsilon_{ij}\sim N\left(0,\sigma^{2}\right)$ and $y_{ij}$ is the squared root of the $j$th CD4 count for the $i$th subject.
The time to event process is specified as:
\begin{eqnarray}
	\displaystyle
	h_{i}(t)=rt^{r-1}\mbox{exp}\left(\beta_{02}+\beta_{12}drug_{i}+\beta_{22}gender_{i}+\beta_{32}prevOI_{i}+\beta_{42}AZT_{i}\right),
	\label{aidsur}
\end{eqnarray}
where $drug_{i}=1$ for ddI, $gender_{i}=1$ for male, $prevOI=1$ (previous opportunistic infection) for AIDS diagnosis and $AZT_{i}=1$ for failure.

	A preliminary study based on fitting the longitudinal process alone by functions, such as \verb|glm| in \verb|R|, indicates an exchangeable structure provides a much better fit compared to an autoregressive structure.
	Thus, the correlation matrix within the longitudinal process is assumed to be exchangeable.  A constant correlation with $\displaystyle\bm R_{(t)(y)}=\rho_{ty}$ and an increasing correlation with $\displaystyle\bm R_{(t)(y)}=\rho_{ty}^{6-j},$ $j=1,2,3,4$ and 5 is considered, as later longitudinal measurements are likely to be have stronger correlation with event time. Two information criteria are applied to gauge the goodness of fit of the proposed models. 
	\begin{eqnarray*}
		\displaystyle
		\mbox{AIC}&=&-2l(\hat{\theta})+2k,\\
		\mbox{BIC}&=&-2l(\hat{\theta})+k\mbox{log}n,\\
	\end{eqnarray*}
	where $\hat{\theta}$ are the parameter estimates of the joint model, $k$ is the number of  parameters being estimated and $n$ is the sample size. 
	
\begin{figure}[H]
		\centering
		\includegraphics[width=\textwidth]{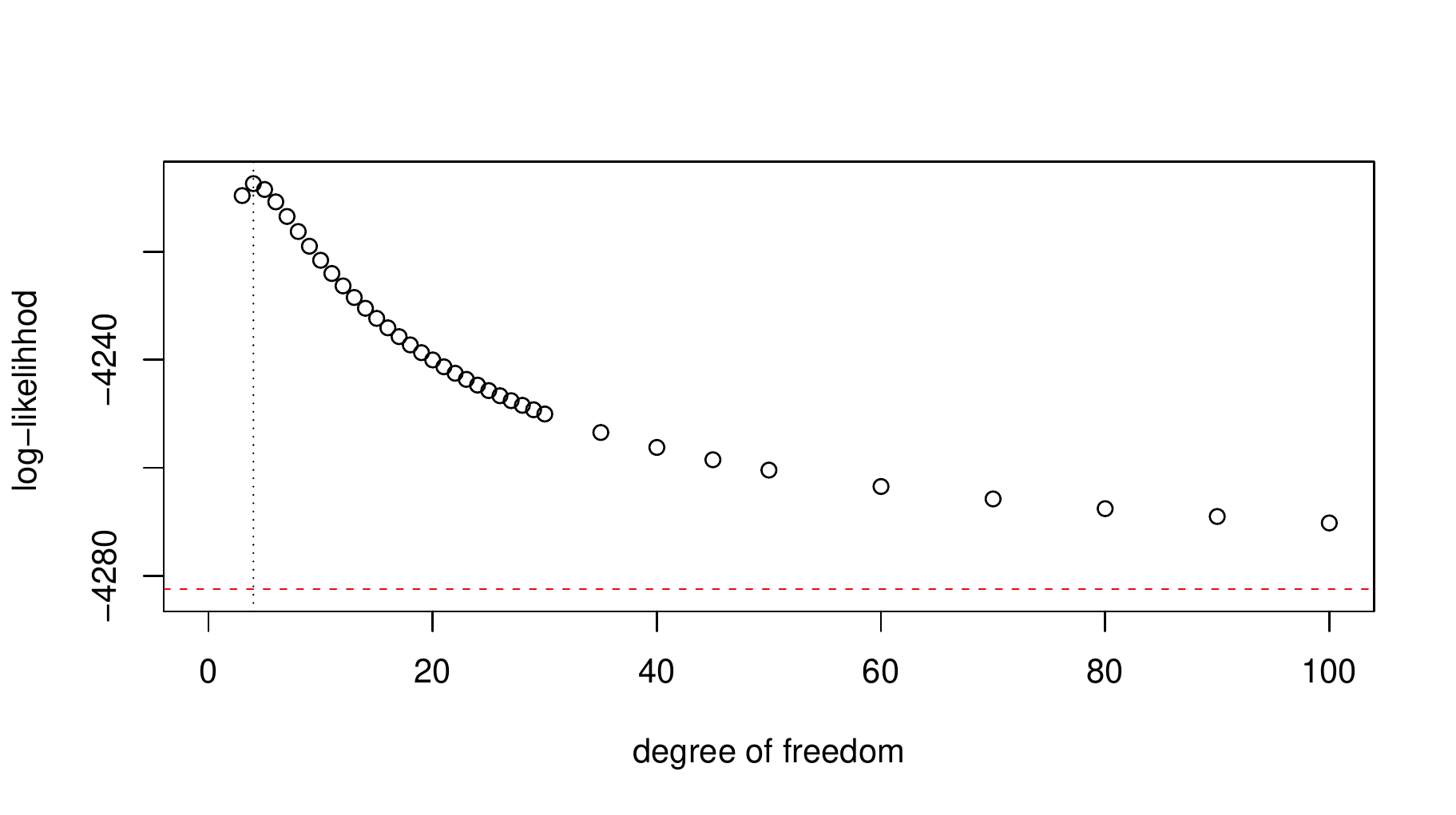}\\
		\caption{\textbf{The log-likelihood values of the fitted multivariate $t$ copula joint model with $\bm R_{(t)(\bm y)}=\rho_{ty}$ and an exchangeable structure for $\bm R_{(\bm y)}$ and the two sub-models as in (\ref{aidslong}) and (\ref{aidsur}) under different degrees of freedom. The red dashed line represents the log-likelihood value of -4282.462 obtained under the Gaussian copula joint model.}}
		\label{loglikivalplot}
	\end{figure}

	We fit the AIDS data by both the multivariate Gaussian and $t$ copula joint models. The multivariate $t$ copula joint model is fitted under various degrees of freedom with $\bm R_{(t)(\bm y)}=\rho_{ty}$ and the corresponding  log-likelihood values are presented in Figure \ref{loglikivalplot}. The optimal fitted model is obtained at $df=4$ with the log-likelihood value of -4209.104 , thus only the results for $df=4$ are presented. As $df$ approaches infinity, the log-likelihood value approaches the value -4282.462 under the multivariate Gaussian copula joint model indicated by the red dashed line. Table \ref{tableaids} summarises the estimation results under both copula joint models.
	
	Under the multivariate Gaussian copula, $\bm R_{(t)(\bm y)}=\bm0$ is equivalent to separately fitting the two sub-models. The fitted results by standard \texttt{R} functions, such as \verb|glm| and \verb|survreg|, for the AIDS data are exactly the same as $\bm R_{(t)(\bm y)}=\bm0$ under the multivariate Gaussian copula joint model in Table \ref{tableaids}, which indicates the proposed model can provide accurate estimates. Under the multivariate $t$ copula joint model with $df=4,$  the $p$-values of 0.242 and 0.182 for the likelihood ratio test and the $z$-test, respectively, suggest $\bm R_{(t)(\bm y)}=\rho_{ty}^{6-j}$ is not significantly better than $\bm R_{(t)(\bm y)}=\bm 0,$ which also matches the conclusion based on AIC and BIC. On the other hand, the $p$-values for the likelihood ratio test and the $z$-test  are almost 0 when comparing $\bm R_{(t)(\bm y)}=\rho_{ty}$ to $\bm R_{(t)(\bm y)}=\bm 0,$ which suggests the former provides a significantly better fit than the latter. 
	Overall, the information criteria indicate the the multivariate $t$ copula joint model with $df=4$ and $\bm R_{(t)(\bm y)}=\rho_{ty}$ provides the best fitting among the six candidates and $\rho_{ty}=0.509$ corresponds to a coefficient of tail dependence of 0.326 between the survival sub-model and each longitudinal measurement under this model.
	
	In practice, it might be reasonable to start with a saturated model for correlation structure in a study with a small
	number of longitudinal measurements (Diggle \textit{et al,} 2008\cite{dig08}), then fit a parsimonious one if a clear pattern can be observed from the fitted saturated correlation structure. Thus, we fit an unstructured $\bm R_{(t)(\bm y)}$ with an exchangeable $\bm R_{(\bm y)}$ for the multivariate $t$ copula with $df=4.$ The fitted values for the components of $\bm R_{(t)(\bm y)}$ are 0.526, 0.585, 0.550, 0.629 and 0.774, which strongly indicate a constant correlation between two sub-models is more appropriate than $\bm R_{(t)(\bm y)}=\rho_{ty}^{6-j},$ although the later longitudinal measurements tend to be slightly more correlated with the event time.

	\begingroup
	\setlength{\tabcolsep}{6pt} 
	\renewcommand{\arraystretch}{1.18} 
	\begin{table}[H]
			\tbl{\textbf{Parameter estimates from the multivariate Gaussian joint model and $t$ copula joint model with $df=4,$ assuming marginals to be (\ref{aidslong}) and (\ref{aidsur}) and $\bm R_{(y)}$ to be an exchangeable structure, for the AIDS data.}}
			{\begin{tabular}{cccccccccccccc}\toprule
					&   \multicolumn{6} {c} {\multirow{2}{*} {Multivariate Gaussian copula joint model}}   &&\multicolumn{6} {c} {\multirow{2}{*} {Multivariate $t$ copula joint model with $df=4$}}
					\\
					\\
					\cline{2-7} \cline{9-14} 
					\\
					&\multicolumn{2} {c}{$\bm R_{(t)(\bm y)}=\rho_{ty}$}&\multicolumn{2} {c}{$\bm R_{(t)(\bm y)}=\rho_{ty}^{6-j}$}	&\multicolumn{2} {c}{$\bm R_{(t)(\bm y)}=\bm0$} &&\multicolumn{2} {c}{$\bm R_{(t)(\bm y)}=\rho_{ty}$}&\multicolumn{2} {c}{$\bm R_{(t)(\bm y)}=\rho_{ty}^{6-j}$}	&\multicolumn{2} {c}{$\bm R_{(t)(\bm y)}=\bm0$} 
					\\
					\\
					\cline{2-7} \cline{9-14} 
					& \multirow{2}{*}{Est.} & \multirow{2}{*}{SE} &  \multirow{2}{*}{Est.}  &  \multirow{2}{*}{SE}  &   \multirow{2}{*}{Est.} & \multirow{2}{*}{SE}  && \multirow{2}{*}{Est.} & \multirow{2}{*}{SE} &\multirow{2}{*}{Est.}  &\multirow{2}{*}{SE}  &\multirow{2}{*}{Est.}  &\multirow{2}{*}{SE} 
					\\
					\\
					\midrule
					$\beta_{01}$           & 10.755  & 0.656   & 10.618  & 0.659  & 10.603  & 0.658  &       &11.194  &0.644 & 11.427    & 0.644  & 11.383 & 0.644
					\\
					$\beta_{11}$          & -0.168  & 0.017     &-0.163   & 0.019 & -0.159   & 0.017   &        & -0.163  & 0.015& -0.161   & 0.016   & -0.155&   0.015
					\\
					$\beta_{21}$          & 0.016   & 0.024     &0.013    & 0.025  &  0.017    & 0.024  &       & 0.018   & 0.021 & 0.012   & 0.021    & 0.019 & 0.021
					\\
					$\beta_{31}$          & -0.464 & 0.646    &-0.303   & 0.649   &  -0.305   & 0.649  &     & -0.701  & 0.644  & -0.685 & 0.645   & -0.676& 0.646
					\\
					$\beta_{41}$          & -4.636  & 0.476    &  -4.631  & 0.476  &-4.623  & 0.476&          & -4.443    & 0.478 & -4.588& 0.475  & -4.564& 0.475
					\\
					$\beta_{51}$         & -0.322  & 0.468   &-0.270  & 0.470  &-0.266  & 0.470  &           & -0.632   & 0.481& -0.673 & 0.478  & -0.661 & 0.478
					\\
					$\beta_{02}$        & -5.320   & 0.386    &-5.366  & 0.387 & -5.362 & 0.387 &           & -5.203   & 0.359& -5.201 & 0.371  & -5.185&   0.371
					\\
					$\beta_{12}$        & 0.236   & 0.135   & 0.207  & 0.147  &  0.213    & 0.146 &             & 0.160    & 0.120  & 0.146 &  0.134  & 0.153 &  0.134
					\\
					$\beta_{22}$        &-0.287 & 0.247     & -0.351  & 0.246&-0.357   & 0.245&              &-0.139    &0.246& -0.292 & 0.238  & -0.301&   0.237
					\\
					$\beta_{32}$       &1.328   & 0.226      & 1.302    & 0.227  &1.300    & 0.227&              &1.419    & 0.225& 1.417   & 0.233   & 1.411  & 0.233
					\\ 
					$\beta_{42}$       & 0.148    & 0.162     &0.159   & 0.163   &  0.158    & 0.163&              &0.198   & 0.151   & 0.172 & 0.147   & 0.162 &  0.147
					\\
					$\rho_{ty}$        & 0.436    & 0.051   & 0.103& 0.186&\textemdash& \textemdash&        &0.509   &0.055  & 0.159 & 0.119 & \textemdash& \textemdash    
					\\
					$\rho_{y}$          & 0.799    & 0.014     & 0.797  & 0.014 &  0.798  & 0.014   &              & 0.837     &0.014  & 0.831 & 0.015  & 0.832    &  0.014
					\\
					$r$                    & 1.355     & 0.094   & 1.417   & 0.094& 1.418     & 0.094 &                &  1.193     &0.083  & 1.307 & 0.084   & 1.306   &  0.084
					\\
					$\sigma$            & 4.372     & 0.126    & 4.353  & 0.125   &   4.353  & 0.125   &           & 4.535     &0.132  & 4.484 & 0.129  & 4.486    &  0.130
					\\
					Loglik&\multicolumn{2} {c}{-4282.462} &\multicolumn{2} {c}{-4310.474}  &\multicolumn{2} {c}{-4310.601}&&\multicolumn{2} {c}{-4207.412}&\multicolumn{2} {c}{-4235.845}&\multicolumn{2} {c}{-4236.529} 
					\\
					AIC    &\multicolumn{2} {c}{8594.923}  &\multicolumn{2} {c}{8650.947}    &\multicolumn{2} {c}{8649.201}   & &\multicolumn{2} {c}{8444.824}&\multicolumn{2} {c}{8501.69}&\multicolumn{2} {c}{8501.057}  
					\\
					BIC    &\multicolumn{2} {c}{8657.118}   &\multicolumn{2} {c}{8713.142}     &\multicolumn{2} {c}{8707.25}   &&\multicolumn{2} {c}{8507.019}&\multicolumn{2} {c}{8563.885} &\multicolumn{2} {c}{8559.106}   
					\\
					\bottomrule
			\end{tabular}}
			\label{tableaids}
	\end{table}
	\endgroup

\begin{figure}[H]
	\centering
	\includegraphics[width=\textwidth]{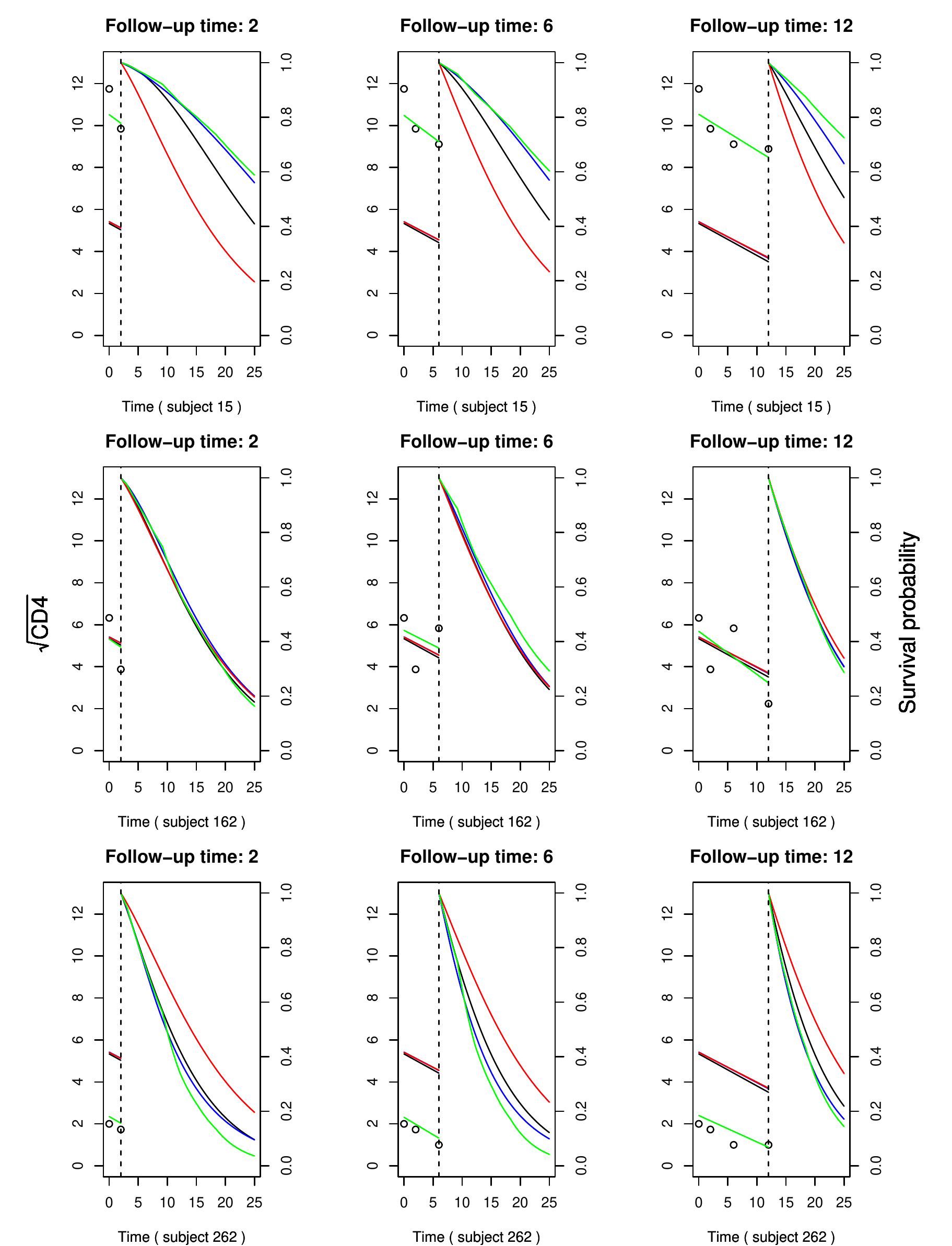}\\
	\caption{\textbf{Dynamic  prediction of survival probabilities and fitted longitudinal trajectories for subjects 15, 162 262 from the AIDS data. The black, red, blue and green lines represent the fitted GJM, SEP, TJM  and the shared random effects joint model with the estimated parameters from Tables \ref{tableaids} and \ref{tablebi}, respectively.}}
	\label{cd4bigplot}
\end{figure}

	In the AIDS data, subjects 15, 162 and 262, are all male treated by ddI with AIDS diagnosis and AZT failure, therefore have the same baseline covariates. While the CD4 levels in subject 15 (censored at $t=12.23$) are greater than 8.889 (over the population trend) for all the measurements at 0, 2, 6 and 12, the CD4 levels in subject 262 (censored at $t=12.23$) are all lower than 2.000 (lower than population trend). On the other hand, subject 162 (censored at $t=15.57$) has CD4 levels around the population trend. Dynamic predictions of survival probabilities, which can be updated as new longitudinal information becomes available, for these three subjects are calculated based on the fitted multivariate Gaussian copula joint models with $\bm R_{(t)(\bm y)}=\rho_{ty}$ (denoted as GJM) and $\bm R_{(t)(\bm y)}=\bm0$ (denoted as SEP) and the multivariate $t$ copula joint model with $\bm R_{(t)(\bm y)}=\rho_{ty}$ and $df=4$ (denoted as TJM) in Table \ref{tableaids}. The results are presented in Figure \ref{cd4bigplot}.

For comparison, a shared random effects joint model  is also fitted and presented by the green line in Figure \ref{cd4bigplot}. It is specified as:
\begin{eqnarray}
	y_{ij}=\beta_{01}+\beta_{11}t_{ij}+\beta_{21}t_{ij}drug_{i}+\beta_{31}gender_{i}+
	\beta_{41}prevOI_{i}+\beta_{51}AZT_{i}+b_{i0}+b_{i1}t+\varepsilon_{ij}
	\label{aidslongbi}
\end{eqnarray}
and
\begin{eqnarray}
	h_{i}(t)=h_{0}(t)\mbox{exp}\left\{\beta_{12}drug_{i}+\beta_{22}gender_{i}+\beta_{32}prevOI_{i}
	+\beta_{42}AZT_{i}+\alpha\left(b_{i0}+b_{i1}t\right)\right\},
	\label{aidsurbi}
\end{eqnarray}
where $\varepsilon_{ij}\sim N(0,\sigma^{2}),$ $(b_{i0},b_{i1})\sim N(0,\bm D) $ and $h_{0}(t)$
is a piecewise-constant baseline function with eight knots equally spaced. The fitted outputs are briefly summarised in Table \ref{tablebi}.

Although there are large differences in the CD4 levels between the three subjects, their fitted longitudinal trajectories under the three copula joint models are identical due to the fact that the marginal models for the longitudinal process only capture the population trend. While the fitted longitudinal trajectories between SEP (red lines), GJM (black line) and TJM (blue line) almost overlap for each subject, we notice some obvious difference in the prediction of survival probabilities between these three models. The SEP  does not take the effects of longitudinal measurements into account, which results in the same prediction of survival probabilities for the three individuals. On the other hand, GJM and TJM calculate the predictions on survival probabilities via the conditional distribution of the event time given the longitudinal process. For example, subject 15 has a CD4 trajectory higher than the population trend, resulting in higher predicted survival probabilities in GJM and TJM than the ones under SEP. Subject 262 displays an opposite trend regardless of sharing the same baseline information as subject 15. Subject 162, on the other hand, shows very little difference among the different models as its trajectory fluctuates around the population trend. This may be due to the fact that the few covariates  included in the longitudinal sub-model are not sufficient to capture all the variations in the longitudinal response. The additional information contained in the error is reflected in the significance of $\rho_{ty},$ which has an impact on the predictions of survival probabilities. If we further compare TJM with GJM, the predicted survival probabilities under TJM provide a larger adjustment from those computed under SEP than GJM when there is a large deviation between the subject's trajectory and the population trend in the longitudinal process. This is expected as the $t$ copula has stronger tail dependence than the  Gaussian copula.
	
	Overall, the difference in regression parameters may not be great between SEP, GJM and TJM,  but the joint model approach is vital for obtaining more accurate predictions in terms of survival probabilities. In addition, the copula joint model provides predictions of survival probabilities which are comparable to the shared random effects joint model  and would be a good alternative to that model when interest is on population inference for the longitudinal process while getting accurate individual predictions of survival probabilities, because it still takes into account any longitudinal effects but without introducing random effects in the models (thus reducing the model complexity).

\begin{table}[H]
	\tbl{Parameter estimates based on the shared random effects joint model specified by (\ref{aidslongbi}) and (\ref{aidsurbi}) for AIDS data.}
	{\begin{tabular}{lcccccccccccc}\toprule
			&$\beta_{01}$& $\beta_{11}$&$\beta_{21}$&$\beta_{31}$ & $\beta_{41}$&$\beta_{51}$  &$\beta_{12}$    &	$\beta_{22}$ &$\beta_{32}$&$\beta_{42}$&$\alpha$ &$\sigma$
			\\
			\midrule
			Est.   & 10.611  & -0.186  & 0.017  & -0.249 & -4.695 &-0.284 & 0.301    & -0.308   & 1.744  & 0.171& -0.246 &1.738
			\\
			SE     & 0.799   & 0.021   & 0.030  &  0.755 & 0.493   &0.474  & 0.167     & 0.452    & 0.395  & 0.201&0.046 &0.048
			\\
			\midrule
			&\multicolumn{4} {c}{Loglik: -4255.669}&\multicolumn{4} {c}{AIC: 8555.338}&\multicolumn{4} {c}{BIC: 8646.557}
			\\
			\bottomrule
	\end{tabular}}
	\label{tablebi}
\end{table}

\section{Discussion}
In this paper, we  optimise the estimating process in the copula joint model, first introduced in Ganjali and Baghfalaki (2015)\cite{gan15}. Computational cost is expected to be reduced while the accuracy of parameter estimation is improved. As well as the multivariate Gaussian copula, we show how a more general $t$ copula can be used as an alternative. The model has flexibility in the choice of  marginal distributions but requires very little additional computation cost compared with the latent random effect joint models. It is capable of providing predictions on survival probabilities at individual level (comparable to a shared random effects joint model) and can be very useful when interest is on the survival event, while also taking into account any effects of longitudinal biomarkers. It is also noticeable that the performance of the multivariate Gaussian copula joint model is very robust as it can provide accurate parameter estimation and offers significant improvement in terms of predicting survival probabilities compared to the survival sub-model alone even when the true joint data is generated by the multivariate $t$ with small $df.$  It also has some computational advantages compared to the multivariate $t$ copula when the longitudinal marginals are normally distributed, which is a typical assumption for continuous biomarkers.  However, the drawbacks of marginal models are not neglectable, especially when compared with random effects models (Diggle et al. 2002\cite{dig02}). Unlike shared or correlated random effects models, it is difficult to place a scientific interpretation on the association parameter $\rho$ in the correlation matrix (Diggle \textit{et al.,} 2008\cite{dig08}). Proposing a reasonable structure for the correlation matrix between the two sub-models is also problematic, especially when there is a large number of longitudinal measurements. Suresh \textit{et al.} (2021a)\cite{sur21a} and (2021b)\cite{sur21b} overcame the issue of finding an appropriate correlation structure between the two sub-models since their approach only requires the specification of two dimensional copulas. However, their approach still falls under the marginal model regime and therefore may be improved by the addition of random effects. A marginal model is only capable of capturing the characteristics of a population trend, thus subjects with the  same covariates are going to exhibit exactly the same tendency for biomarkers, except differences in measurement errors. A standard deviation of more than 4 for the random errors under the copula  joint model in the presentation of the fitted models for the AIDS data in Table \ref{tableaids} is quite large compared to the magnitude of the longitudinal measurements, indicating there is still a certain amount of individual information contained in the random error. The shared random effects joint model can effectively reduce the standard deviation for random errors to 1.738 (Table \ref{tablebi}), as subject deviation from the population is accounted for by random effects. This is consistent with the fitted curves in the longitudinal process in Figure \ref{cd4bigplot}, where the green lines capture the observed individual measurements much better than the red, black and blue ones.  

Although Rizopoulos \textit{et al.} (2008a\cite{riz08a} and 2008b\cite{riz08b}) and Malehi \textit{et al.,} (2015)\cite{mal15} extended the flexibility of dependency between event time and longitudinal processes by using copula, their models are  still under the framework of a conditional independence assumption, which can be difficult to verify for real data.  We would like to explore the possibility of combining the advantages of a copula joint model (introducing non linear correlation or even checking the assumption of conditional independence between the two processes) and the shared or correlated random effects joint model (straightforward for subject-specific study). To achieve this, we are currently extending the estimation approach proposed in this paper to fit a copula joint model with shared random effects and the results seem very promising and will be reported elsewhere. The key step of this new model is to use a copula to link the conditional distributions of the longitudinal and survival processes given random effects. 

Other further extensions could include exploring the multivariate skew normal or $t$ copulas (Demarta and McNeil, 2005\cite{dem05}), which will allow the joint model to reflect an asymmetric association at the two tails.

\section*{Software}
The R code used in this work is available at https://github.com/zhangzili0916/jointmodel-copula-approach on GitHub.

\section*{Disclosure statement}

No potential conflict of interest was reported by the authors.

\section*{Appendix A}

Suppose the true event times are observed for all the subjects in a group with sample size $n$. Taking $\bm Z=\bm Z\left(\cdot\right)$  for simplicity, the log-likelihood is given by:

\begin{eqnarray*}
	\displaystyle
	l(\bm\theta)&=&Const-\sum_{i=1}^{n}m_{i}\mbox{log}\sigma-\frac{1}{2}\sum_{i=1}^{n}\mbox{log}|	\bm R(\bm\rho)_{(t_{i},\bm y_{i})}|-\frac{1}{2}\sum_{i=1}^{n}\bm Z_{i}^{'}	\bm R(\bm\rho)_{(t_{i},\bm y_{i})}^{-1}\bm Z_{i}\\
	\displaystyle
	&&+\sum_{i=1}^{n}\mbox{log}f_{T_{i}}(t_{i};\bm x_{i2}^{'}\bm\beta_{2},r)-\sum_{i=1}^{n}\mbox{log}\phi\left(Z_{t_{i}}\right),
\end{eqnarray*}

Let
\begin{eqnarray*}
	\displaystyle
	\bm R(\bm\rho)_{(t_{i},\bm y_{i})}^{-1}=\left(
	\begin{array}{ll}
		\displaystyle
		A_{i}& \bm B_{i}^{'}
		\\
		\displaystyle
		\bm B_{i}&\bm D_{i}
	\end{array}
	\right)
\end{eqnarray*}
where (take $\bm R=\bm R(\cdot)$ for simplicity) \[A_{i}=\left(1-\bm R_{(t_{i})(\bm y_{i})}\bm R_{(\bm y_{i})}^{-1}\bm R_{(\bm y_{i})(t_{i})}\right)^{-1},\]
\[\bm B_{i}=-\left(1-\bm R_{(t_{i})(\bm y_{i})}\bm R_{(\bm y_{i})}^{-1}\bm R_{(\bm y_{i})((t_{i})}\right)^{-1}\bm R_{(\bm y_{i})}^{-1}\bm R_{(\bm y_{i})(t_{i})}\]
and
\[\bm D_{i}=\bm R_{(\bm y_{i})}^{-1}+\bm R_{(\bm y_{i})}^{-1}\bm R_{(\bm y_{i})(t_{i})}\left(1-\bm R_{(t_{i})(\bm y_{i})}\bm R_{(\bm y_{i})}^{-1}\bm R_{(\bm y_{i})(t_{i})}\right)^{-1}\bm R_{(t_{i})(\bm y_{i})}\bm R_{(\bm y_{i})}^{-1}.\]
Thus we can rewrite the log-likelihood as:
\begin{eqnarray}
	\displaystyle
	l(\bm\theta)&=&Const-\sum_{i=1}^{n}m_{i}\mbox{log}\sigma-\frac{1}{2}\sum_{i=1}^{n}\mbox{log}|\bm R_{(t_{i},\bm y_{i})}|-\frac{1}{2}\sum_{i=1}^{n}\left(Z_{t_{i}}\mbox{ }\bm Z_{\bm y_{i}}^{'}\right)\left(
	\begin{array}{ll}
		\displaystyle
		A_{i}& \bm B_{i}^{'}
		\\
		\displaystyle
		\bm B_{i}&\bm D_{i}
	\end{array}
	\right)
	\left(
	\begin{array}{l}
		\displaystyle
		Z_{t_{i}}
		\\
		\displaystyle
		\bm Z_{\bm y_{i}}
	\end{array}
	\right) \nonumber
	\\
	\displaystyle
	&&+\sum_{i=1}^{n}\mbox{log}f_{T_{i}}(t_{i};\bm x_{i2}^{'}\bm\beta_{2},r)-\sum_{i=1}^{n}\mbox{log}\phi\left(Z_{t_{i}}\right) \nonumber
	\\
	&=&Const-\sum_{i=1}^{n}m_{i}\mbox{log}\sigma-\frac{1}{2}\sum_{i=1}^{n}\mbox{log}|\bm R_{(t_{i},\bm y_{i})}|-\frac{1}{2}\sum_{i=1}^{n}\left(A_{i}Z_{t_{i}}^{2}+\bm Z_{\bm y_{i}}^{'}\bm B_{i} Z_{t_{i}}+Z_{t_{i}}\bm B_{i}^{'}\bm Z_{\bm y_{i}}+\bm Z_{\bm y_{i}}^{'}\bm D_{i}\bm Z_{\bm y_{i}}\right) \nonumber
	\\
	&&+\sum_{i=1}^{n}\mbox{log}f_{T_{i}}(t_{i};\bm x_{i2}^{'}\bm\beta_{2},r)-\sum_{i=1}^{n}\mbox{log}\phi\left(Z_{t_{i}}\right)
	\label{4}
\end{eqnarray}
The score equation of $\bm\beta_{1}$ is given as follow:
\begin{eqnarray*}
	\displaystyle
	\frac{\partial l(\bm\theta)}{\partial\bm\beta_{1}}&=&
	\partial\left\{-\frac{1}{2}\sum_{i=1}^{n}\left(\bm Z_{\bm y_{i}}^{'}\bm B_{i} Z_{t_{i}}+Z_{t_{i}}\bm B_{i}^{'}\bm Z_{\bm y_{i}}+\bm Z_{\bm y_{i}}^{'}\bm D_{i}\bm Z_{\bm y_{i}}\right)\right\}/\partial\bm\beta_{1}
	\\
	&=&-\frac{1}{2\sigma}\sum_{i=1}^{n}\left(-2\bm X_{i1}^{'}\bm B_{i} Z_{t_{i}}-2\bm X_{i1}^{'}\bm D_{i}\bm Z_{\bm y_{i}}\right)
	\\
	&=&\bm X_{1}^{'}\bunderline{\bm B} \bm Z_{t}+\bm X_{1}^{'}\bunderline{\bm D}\bm\left(\bm Y-\bm X_{1}\bm\beta_{1}\right)/\sigma
	\\
	&=&\bm 0
\end{eqnarray*}
Solving the equation gives
\begin{eqnarray}
	\displaystyle
	\hat{\bm\beta}_{1}(\bm\beta_{2},\bm\alpha,r,\sigma)&=&\left(\bm X_{1}^{'}\bunderline{\bm D}\bm X_{1}\right)^{-1}\bm X_{1}^{'}\left(\sigma\bunderline{\bm B}\bm Z_{t}+\underline{\bm D}\bm Y\right),
	\label{5}
\end{eqnarray}

where $\bm X_{1}=\left(\bm X_{11}^{'},...,\bm X_{n1}^{'}\right)^{'},$ $\bm Y=\left(\bm y_{1}^{'},...,\bm y_{n}^{'}\right)^{'},$ $\bm Z_{t}=\left(Z_{t_{1}},...,Z_{t_{n}}\right)^{'},$ 
$\bunderline{\bm B}=\left(
\begin{array}{lll}
	\displaystyle
	\bm B_{1}& &
	\\
	\displaystyle
	& \ddots&
	\\
	&&\bm B_{n}
\end{array}
\right)_{(\sum_{i=1}^{n}m_{i})\times n}$ and
$\bunderline{\bm D}=\left(
\begin{array}{lll}
	\displaystyle
	\bm D_{1}& &
	\\
	\displaystyle
	& \ddots&
	\\
	&&\bm D_{n}
\end{array}
\right)_{(\sum_{i=1}^{n}m_{i})\times (\sum_{i=1}^{n}m_{i})}.$

Substituting (\ref{5}) back in the log-likelihood function (\ref{4}) then differentiating with respect to $\sigma$ leads to

\begin{eqnarray}
	\displaystyle
	\frac{\partial l(\bm\theta)}{\partial\sigma}&=&\partial\left\{-\sum_{i=1}^{n}m_{i}\mbox{log}\sigma-\frac{1}{2}\sum_{i=1}^{n}\left(\frac{2}{\sigma}Z_{t_{i}}\bm B_{i}^{'}\left(\bm y_{i}-\bm X_{i1}\hat{\bm\beta}_{1}\right)+\frac{1}{\sigma^{2}}\left(\bm y_{i}-\bm X_{i1}\hat{\bm\beta}_{1}\right)^{'}\bm D_{i}\left(\bm y_{i}-\bm X_{i1}\hat{\bm\beta}_{1}\right)\right)\right\}/\partial\sigma  \nonumber
	\\
	&=&\partial\left\{-\sum_{i=1}^{n}m_{i}\mbox{log}\sigma-\frac{1}{\sigma}\bm Z_{t}^{'}\bunderline{\bm B}^{'}\left(\bm Y-\bm X_{1}\hat{\bm\beta}_{1}\right)-\frac{1}{2\sigma^{2}}\left(\bm Y-\bm X_{1}\hat{\bm\beta}_{1}\right)^{'}\bunderline{\bm D}\left(\bm Y-\bm X_{1}\hat{\bm\beta}_{1}\right)\right\}/\partial\sigma  \nonumber
	\\
	&=&0
	\label{6}
\end{eqnarray}
part of the numerator of  equation (\ref{6}) can be simplified as:
\begin{eqnarray}
	&&-\frac{1}{\sigma}\bm Z_{t}^{'}\bunderline{\bm B}^{'}\left(\bm Y-\bm X_{1}\hat{\bm\beta}_{1}\right)-\frac{1}{2\sigma^{2}}\left(\bm Y-\bm X_{1}\hat{\bm\beta}_{1}\right)^{'}\bunderline{\bm D}\left(\bm Y-\bm X_{1}\hat{\bm\beta}_{1}\right) \nonumber
	\\
	&=&-\frac{1}{\sigma}\bm Z_{t}^{'}\bunderline{\bm B}^{'}\left(\bm Y-\bm X_{1}\left(\bm X_{1}^{'}\bunderline{\bm D}\bm X_{1}\right)^{-1}\bm X_{1}^{'}\left(\sigma\bunderline{\bm B}\bm Z_{t}+\bunderline{\bm D}\bm Y\right)\right)  \nonumber
	\\
	&&-\frac{1}{2\sigma^{2}}\left(\bm Y-\bm X_{1}\left(\bm X_{1}^{'}\bunderline{\bm D}\bm X_{1}\right)^{-1}\bm X_{1}^{'}\left(\sigma\bunderline{\bm B}\bm Z_{t}+\bunderline{\bm D}\bm Y\right)\right)^{'}\bunderline{\bm D}\left(\bm Y-\bm X_{1}\left(\bm X_{1}^{'}\bunderline{\bm D}\bm X_{1}\right)^{-1}\bm X_{1}^{'}\left(\sigma\bunderline{\bm B}\bm Z_{t}+\bunderline{\bm D}\bm Y\right)\right) \nonumber
	\\
	&=&-\frac{1}{2\sigma^{2}}\bm Y^{'}\left(\bunderline{\bm D}-\bunderline{\bm D}\bm X_{1}\left(\bm X_{1}^{'}\bunderline{\bm D}\bm X_{1}\right)^{-1}\bm X_{1}^{'}\bunderline{\bm D}\right)\bm Y-\frac{1}{\sigma}\bm Z_{t}^{'}\left(\bunderline{\bm B}^{'}-\bunderline{\bm B}^{'}\bm X_{1}\left(\bm X_{1}^{'}\bunderline{\bm D}\bm X_{1}\right)^{-1}\bm X_{1}^{'}\bunderline{\bm D}\right)\bm Y    \nonumber
	\\
	&&+\frac{1}{2}\bm Z_{t}^{'}\bunderline{\bm B}^{'}\bm X_{1}\left(\bm X_{1}^{'}\bunderline{\bm D}\bm X_{1}\right)^{-1}\bm X_{1}^{'}\bunderline{\bm B}\bm Z_{t}   \nonumber
	\\
	&=&\frac{1}{\sigma^{2}} G+\frac{1}{\sigma} H+Const,
	\label{7}
\end{eqnarray}
where  $\displaystyle G=-\frac{1}{2}\bm Y^{'}\left(\bunderline{\bm D}-\bunderline{\bm D}\bm X_{1}\left(\bm X_{1}^{'}\bunderline{\bm D}\bm X_{1}\right)^{-1}\bm X_{1}^{'}\bunderline{\bm D}\right)\bm Y,$  (Note: $G\leq0$ since $\displaystyle G=-\frac{1}{2}(\bm Y-\bm X_{1}\tilde{\bm\beta})^{'}\bunderline{\bm D}(\bm Y-\bm X_{1}\tilde{\bm\beta})$ with $\tilde{\bm\beta}=(\bm X_{1}^{'}\bunderline{\bm D}\bm X_{1})^{-1}\bm X_{1}^{'}\bunderline{\bm D}\bm Y),$ 
$H=-\bm Z_{t}^{'}\left(\bunderline{\bm B}^{'}-\bunderline{\bm B}^{'}\bm X_{1}\left(\bm X_{1}^{'}\bunderline{\bm D}\bm X_{1}\right)^{-1}\bm X_{1}^{'}\bunderline{\bm D}\right)\bm Y$  and $Const=\frac{1}{2}\bm Z_{t}^{'}\bunderline{\bm B}^{'}\bm X_{1}\left(\bm X_{1}^{'}\bunderline{\bm D}\bm X_{1}\right)^{-1}\bm X_{1}^{'}\bunderline{\bm B}\bm Z_{t}.$

Substituting (\ref{7}) back to (\ref{6}) gives:
\begin{eqnarray}
	(\sum_{i=1}^{n}m_{i})\sigma^{2}+H\sigma+2G=0
	\label{8}
\end{eqnarray}

The MLE of $\sigma$ is the positive root of equation (\ref{8}),
i.e.,
\begin{eqnarray}
	\hat{\sigma}(\bm\beta_{2},\bm\rho,r)=\frac{-H+\sqrt{H^2-8(\sum_{i=1}^{n}m_{i})G}}{2(\sum_{i=1}^{n}m_{i})}
	\label{9}
\end{eqnarray}

Assuming the survival data follows a Weibull distribution, $\displaystyle\frac{\partial l(\bm\theta)}{\partial\bm\beta_{2}}=\bm0$, $\displaystyle\frac{\partial l(\bm\theta)}{\partial r}=0$ and $\displaystyle\frac{\partial l(\bm\theta)}{\partial\bm\rho}=\bm0$ do not have explicitly solutions, thus numerical optimisation technique is required.

\section*{Appendix B}
Tables 6 to 7 present the fitted outputs produced by the five candidate models under dropout rates 60\% and 80\% in simulation study 1 while tables 8 to 9 present the fitted outputs produced by the four candidate models under dropout rates  60\% and 80\% in simulation study 2. For detailed discussion, please refer to Section 3.2. 

\begin{landscape}
	\vspace*{\fill}
	\begingroup
	\setlength{\tabcolsep}{6pt} 
	\renewcommand{\arraystretch}{1.18} 
	\begin{table}[H]
			\tbl{\textbf{{\scriptsize Data are generated by the multivariate Gaussian copula  joint model with 60\% dropout (67\% censoring) rate and estimate the parameters by the multivariate Gaussian copula  joint model and multivariate $t$ copula joint model with 3, 4, 40 and 100 degree of freedom when the marginal distributions and $\bm R(\bm\rho)_{(t,\bm y)}$ are correctly specified.}}}
			{\begin{tabular}{lccccccccccccccccc}
					\toprule
					True value &$\beta_{01}$&$\beta_{11}$&$\beta_{21}$&$\beta_{31}$&$\beta_{41}$&$\beta_{51}$&$\beta_{12}$&$\beta_{22}$&$\beta_{32}$&$\beta_{42}$&$\beta_{52}$&$r$&$\sigma$&$\rho_{ty}$  &$\rho_{y}$
					\\
					& 5         & 1         & 2                 & 1             &  -2         &   -1        & -5        &-4      & -2       & 2   &1    & 2       &  3       &  0.6         & 0.4      
					\\
					\hline
					Gaussian copula &                   &                   &                    &                   &                   &                   &                    &                    &                     &                   &
					\\
					\cdashline{2-16}
					Est.        &5.011  & 0.999  & 2.002  & 0.987 &-2.002 & -1.003 & -5.130 & -4.172 &-2.053 &2.049 &1.044 & 2.053 & 2.974  & 0.603  & 0.389
					\\
					SE         & 0.287 & 0.028  &  0.031 &  0.316 & 0.363 & 0.409  & 0.485  & 0.452  & 0.258 & 0.275 &0.291 & 0.181 & 0.096  & 0.039  & 0.041 
					\\
					SD         & 0.279 & 0.028 & 0.032  & 0.324  &0.347  &0.409   &0.495   &0.475   &0.276  &0.277  &0.303 &0.185  &0.096   & 0.039  &0.042  
					\\
					RMSE    & 0.279 & 0.028  & 0.032 &  0.324 &0.347  &0.409   &0.511    & 0.505  & 0.280 & 0.281 &0.306 &0.192  & 0.099  & 0.040  & 0.043
					\\
					CP	      & 0.954 & 0.954 & 0.940  & 0.936  & 0.956 & 0.954  &0.942   &0.970   & 0.934 &0.942 &0.930 & 0.944 & 0.940  &0.942   & 0.940   
					\\
					\cdashline{1-16}
					$t$ copula ($df=3$)  & &  &   &   &    &   &  &  &  &  &
					\\
					\cdashline{2-16}
					Est.     &5.036   & 0.998 &2.001  &0.983   &-1.992 &-1.005 &-4.662 &-3.736 &-1.859 & 1.846  &0.936 & 1.855  & 3.356  &0.577  &0.357
					\\
					SE       &0.289   & 0.028 &  0.031&  0.318 & 0.365 & 0.411  & 0.420  & 0.370  & 0.230 & 0.244 &0.263 &0.156   & 0.111   & 0.047 &0.048
					\\
					SD       & 0.304  & 0.030 & 0.035 &  0.361 & 0.381 & 0.446 & 0.465  & 0.423  & 0.267 & 0.268 &0.292 & 0.172  & 0.120  & 0.043 & 0.043 
					\\
					RMSE  & 0.306  & 0.030 &0.035  & 0.362  & 0.380 & 0.446 &0.575  & 0.499  & 0.302 & 0.308 &0.299 & 0.225 & 0.376  & 0.049 & 0.061
					\\
					CP	    & 0.944  & 0.934 & 0.918 & 0.922  & 0.938 & 0.940 &0.812   &0.812   &0.842   & 0.870 &0.918  & 0.788 & 0.116  & 0.954 & 0.886
					\\
					\cdashline{1-16}
					$t$ copula ($df=4$)  & &  &   &   &    &   &  &  &  &  &
					\\
					\cdashline{2-16}
					Est.     &5.028   & 0.998 &2.001  &0.983  &-1.995 &-1.004 &-4.827 &-3.893 &-1.929 & 1.918   &0.973 & 1.925  & 3.222  &0.582 &0.364
					\\
					SE       &0.288   & 0.028 &  0.031&  0.317 & 0.364 & 0.409 & 0.441  & 0.394  & 0.239 & 0.253 &0.272 &0.164   & 0.108  & 0.045 &0.047
					\\
					SD      & 0.295  & 0.029 & 0.034 &  0.349 & 0.369 & 0.432 & 0.477  & 0.439 & 0.273  & 0.272 &0.297  & 0.177  & 0.111   & 0.042 & 0.042 
					\\
					RMSE  & 0.296 & 0.029  &0.034  & 0.349 & 0.368  & 0.431 &0.506  & 0.451  & 0.282  & 0.284 &0.298  & 0.192 & 0.248  & 0.046 & 0.056
					\\
					CP	    & 0.950 & 0.938  & 0.930 & 0.920 & 0.944  & 0.950 &0.888  &0.872   &0.898   & 0.896 &0.928  & 0.886 & 0.466 & 0.960 & 0.902
					\\
					\cdashline{1-16}
					$t$ copula ($df=40$)   & &  &   &   &    &   &  &  &  &  &
					\\
					\cdashline{2-16}
					Est.    & 5.012   &0.998  & 2.002   &0.988  & -2.005 & -1.007 &-5.127 & -4.169 & -2.052 &2.050 &1.043   &2.051   &2.979   &0.602  &0.388
					\\
					SE      & 0.287  & 0.028 & 0.031    &  0.316 &  0.362 & 0.409  & 0.484 & 0.450  & 0.258  & 0.274 &0.291   & 0.181  & 0.097  & 0.040 &0.042 
					\\
					SD      & 0.278  & 0.028 &  0.032  &  0.325 & 0.346  &  0.407 & 0.496 & 0.472  & 0.276   & 0.278 & 0.302 & 0.185  & 0.096  & 0.039 & 0.041
					\\
					RMSE & 0.278  & 0.028 & 0.032   & 0.325  & 0.345  & 0.407  & 0.511  & 0.501  & 0.281   & 0.282  &0.305  & 0.192  & 0.098  & 0.040 & 0.043
					\\
					CP	    & 0.954 & 0.958 & 0.946   & 0.934  & 0.956  & 0.954  &0.944  &0.970   & 0.934  & 0.942  &0.938  & 0.944  & 0.944  & 0.942 & 0.936  
					\\
					\cdashline{1-16}
					$t$ copula ($df=100$)   & &  &   &   &    &   &  &  &  &  &
					\\
					\cdashline{2-16}
					Est.    &5.012    &0.998  & 2.002  &0.987  & -2.005 & -1.006  &-5.130 & -4.171 & -2.054&2.052   &1.044  &2.052  &2.976   &0.603    &0.389
					\\
					SE      & 0.287  & 0.028 &  0.031 &  0.316 &  0.362 & 0.409   & 0.484  & 0.451  & 0.258  & 0.275  &0.291  & 0.181  & 0.096  & 0.039   & 0.041
					\\
					SD      & 0.278  & 0.028 &  0.032&  0.324 & 0.347  &  0.407  & 0.495  & 0.473  & 0.275  & 0.278   &0.303 & 0.185 & 0.096   &  0.039 & 0.042 
					\\
					RMSE & 0.278  & 0.028 & 0.032  &0.325   &0.347   & 0.407   & 0.511  & 0.503  & 0.280  & 0.283  & 0.305 & 0.192 &0.099   &  0.039  & 0.043  
					\\
					CP	    & 0.954 & 0.954 & 0.940  & 0.932  & 0.956  & 0.962   &0.944  &0.970   &0.934   & 0.942  &0.934  & 0.944 & 0.940 & 0.944   & 0.940
					\\
					\hline
			\end{tabular}}
			\label{simGau2}
	\end{table}
	\endgroup
	\vspace*{\fill}
\end{landscape}

\begin{landscape}
	\vspace*{\fill}
	\begingroup
	\setlength{\tabcolsep}{6pt} 
	\renewcommand{\arraystretch}{1.18} 
	\begin{table}[H]
			\tbl{\textbf{{\scriptsize Data are generated by the multivariate Gaussian copula  joint model with 80\% dropout (75\% censoring) rate and estimate the parameters by the multivariate Gaussian copula  joint model and multivariate $t$ copula joint model with 3, 4, 40 and 100 degree of freedom when the marginal distributions and $\bm R(\bm\rho)_{(t,\bm y)}$ are correctly specified.}}}
			{\begin{tabular}{lccccccccccccccccc}
					\toprule
					True value &$\beta_{01}$&$\beta_{11}$&$\beta_{21}$&$\beta_{31}$&$\beta_{41}$&$\beta_{51}$&$\beta_{12}$&$\beta_{22}$&$\beta_{32}$&$\beta_{42}$&$\beta_{52}$&$r$&$\sigma$&$\rho_{ty}$  &$\rho_{y}$
					\\
					& 5         & 1         & 2                 & 1             &  -2         &   -1        & -5        &-4      & -2       & 2   &1    & 2       &  3       &  0.6         & 0.4      
					\\
					\hline
					Gaussian copula &                   &                   &                    &                   &                   &                   &                    &                    &                     &                   &
					\\
					\cdashline{2-16}
					Est.        &5.016  & 0.997  & 2.003  & 0.979 &-1.998 & -1.001  & -5.149  & -4.191 &-2.070 &2.050  &1.041  & 2.065 & 2.975  & 0.608  & 0.390
					\\
					SE         & 0.298 & 0.035  &  0.040 &  0.330 & 0.379 & 0.429  & 0.534  & 0.549  & 0.298 & 0.313 &0.337  & 0.204 & 0.102  & 0.044  & 0.046 
					\\
					SD        & 0.288  & 0.034 & 0.040  & 0.338  &0.366  &0.436   &0.561   &0.539   &0.321   &0.322  &0.356  &0.219  &0.105   & 0.045  &0.044  
					\\
					RMSE    & 0.288 & 0.034  & 0.040 &  0.338 &0.365  &0.435   &0.580   & 0.572  & 0.328 & 0.326 &0.358  &0.228  & 0.107  & 0.046  & 0.046
					\\
					CP	      & 0.952 & 0.954 & 0.956  & 0.942  & 0.954 & 0.948  &0.936   &0.970    & 0.936 &0.950  &0.944  & 0.930 & 0.934  &0.928   & 0.952   
					\\
					\cdashline{1-16}
					$t$ copula ($df=3$)  & &  &   &   &    &   &  &  &  &  &
					\\
					\cdashline{2-16}
					Est.     &5.029   & 0.997 &2.002  &0.973   &-1.985 &-0.994 &-4.742 &-3.822 &-1.904 & 1.876  &0.950 & 1.894  & 3.292  &0.575  &0.351
					\\
					SE       &0.299   & 0.035 &  0.040&  0.331 & 0.381 & 0.429  & 0.468  & 0.453  & 0.270 & 0.282 &0.312 &0.178   & 0.116   & 0.053 &0.053
					\\
					SD       & 0.310  & 0.038 & 0.045 &  0.368 & 0.399 & 0.470 & 0.532  & 0.489  & 0.315 & 0.310 &0.348 & 0.206  & 0.125  & 0.049 & 0.045 
					\\
					RMSE  & 0.311  & 0.038 &0.045  & 0.368  & 0.399 & 0.470 &0.590    & 0.520  & 0.329 & 0.333 &0.351 & 0.231  & 0.317   & 0.055 & 0.067
					\\
					CP	    & 0.934  & 0.928 & 0.912 & 0.928  & 0.934 & 0.932 &0.842   &0.862   &0.888   & 0.900 &0.912  & 0.836 & 0.296  & 0.962 & 0.876
					\\
					\cdashline{1-16}
					$t$ copula ($df=4$)  & &  &   &   &    &   &  &  &  &  &
					\\
					\cdashline{2-16}
					Est.     &5.025   & 0.997 &2.003  &0.974  &-1.988 &-0.994 &-4.888 &-3.958 &-1.966 & 1.938   &0.981 & 1.956  & 3.180  &0.583 &0.359
					\\
					SE       &0.298   & 0.035 &  0.040&  0.330 & 0.380 & 0.428 & 0.489  & 0.480  & 0.279 & 0.292 &0.321 &0.186   & 0.112  & 0.052 &0.052
					\\
					SD      & 0.302  & 0.037 & 0.044 &  0.357 & 0.387 & 0.457 & 0.543  & 0.505 & 0.320  & 0.315 &0.352  & 0.211  & 0.117   & 0.048 & 0.045 
					\\
					RMSE  & 0.303 & 0.037  &0.044  & 0.357 & 0.386  & 0.456 &0.554  & 0.506  & 0.321  & 0.320 &0.352  & 0.215 & 0.215  & 0.051 & 0.060
					\\
					CP	    & 0.944 & 0.940  & 0.926 & 0.934 & 0.938  & 0.940 &0.894  &0.922   &0.914   & 0.930 &0.920  & 0.896 & 0.638 & 0.952 & 0.890
					\\
					\cdashline{1-16}
					$t$ copula ($df=40$)   & &  &   &   &    &   &  &  &  &  &
					\\
					\cdashline{2-16}
					Est.    & 5.017   &0.997  & 2.004   &0.979   & -1.999 & -1.001 &-5.147  & -4.189 & -2.069 &2.049 &1.040   &2.064   &2.980   &0.607  &0.388
					\\
					SE      & 0.298  & 0.035 & 0.040   &  0.330 &  0.379 & 0.428  & 0.532 & 0.544  & 0.297   & 0.312 &0.336   & 0.204  & 0.103  & 0.045 &0.047 
					\\
					SD      & 0.288  & 0.034 &  0.040  &  0.339 & 0.366  &  0.436 & 0.561 & 0.537  & 0.322    & 0.322 & 0.356 & 0.219  & 0.105  & 0.045 & 0.045
					\\
					RMSE & 0.288  & 0.034 & 0.040   & 0.339  & 0.366  & 0.436  & 0.580  & 0.569  & 0.329   & 0.325 &0.358  & 0.228  & 0.107  & 0.046 & 0.046
					\\
					CP	    & 0.958 & 0.956 & 0.954   & 0.942  & 0.952  & 0.940  &0.938   &0.968    & 0.938  & 0.944  &0.944  & 0.928  & 0.932  & 0.936 & 0.950  
					\\
					\cdashline{1-16}
					$t$ copula ($df=100$)   & &  &   &   &    &   &  &  &  &  &
					\\
					\cdashline{2-16}
					Est.    &5.017    &0.997  & 2.004  &0.980  & -1.998 & -1.003  &-5.150 & -4.190 & -2.071 &2.050   &1.042  &2.066  &2.976   &0.608    &0.389
					\\
					SE      & 0.298  & 0.035 &  0.040 &  0.330 &  0.379 & 0.428   & 0.534 & 0.547  & 0.298  & 0.312  &0.337  & 0.204 & 0.103  & 0.044   & 0.046
					\\
					SD      & 0.288  & 0.034 &  0.040&  0.339 & 0.366  &  0.435  & 0.562  & 0.540 & 0.322  & 0.323  &0.355  & 0.219 & 0.105   &  0.045 & 0.045
					\\
					RMSE & 0.289  & 0.034 & 0.040  &0.339   &0.366   & 0.435   & 0.581  & 0.572  & 0.329  & 0.326  & 0.358 & 0.228 &0.107   &  0.046  & 0.046  
					\\
					CP	    & 0.952 & 0.958 & 0.954  & 0.944  & 0.956  & 0.946   &0.934   &0.970   &0.940   & 0.946  &0.946  & 0.932 & 0.932 & 0.926   & 0.954
					\\
					\hline
			\end{tabular}}
			\label{simGau3}
	\end{table}
	\endgroup
	\vspace*{\fill}
\end{landscape}

\begin{landscape}
	\vspace*{\fill}
	\begingroup
	\setlength{\tabcolsep}{6pt} 
	\renewcommand{\arraystretch}{1.18} 
	\begin{table}[H]
			\tbl{\textbf{{\scriptsize Data are generated by the multivariate $t$ copula joint model with 4 degree of freedom and 60\% dropout (67\% censoring) rate and estimate the parameters by the multivariate Gaussian copula and multivariate $t$ copula joint model with 3, 4 and 5 degree of freedom when the marginal distributions and $\bm R(\bm\rho)_{(t,\bm y)}$ are correctly specified.}}}
			{\begin{tabular}{lccccccccccccccccc}
					\toprule
					True value &$\beta_{01}$&$\beta_{11}$&$\beta_{21}$&$\beta_{31}$&$\beta_{41}$&$\beta_{51}$&$\beta_{12}$&$\beta_{22}$&$\beta_{32}$&$\beta_{42}$&$\beta_{52}$&$r$&$\sigma$&$\rho_{ty}$  &$\rho_{y}$
					\\
					& 5         & 1         & 2                 & 1             &  -2         &   -1        & -5        &-4      & -2       & 2   &1    & 2       &  3       &  0.6         & 0.4      
					\\
					\hline
					$t$ copula ($df=4$) &                   &                   &                    &                   &                   &                   &                    &                    &                     &                   &
					\\
					\cdashline{2-16}
					Est.       &5.009  & 0.999  & 2.005  & 0.991  &-2.004  & -0.999 & -5.140 & -4.131 &-2.059 &2.068  &1.055 & 2.054 & 2.981  & 0.605  & 0.395
					\\
					SE         & 0.266 & 0.025  &  0.028 &  0.292 & 0.335  & 0.377   & 0.448  & 0.397  & 0.236 & 0.252 &0.266 & 0.167  & 0.101  & 0.043  & 0.046
					\\
					SD         & 0.254 & 0.025  & 0.029  & 0.290  &0.335  &0.378    &0.446   &0.412   &0.236   &0.266  &0.267 &0.165   &0.093  & 0.045  &0.044  
					\\
					RMSE    & 0.254 & 0.025  & 0.029  &  0.290 &0.335   &0.378    &0.467   & 0.432  & 0.243 & 0.274  &0.273 &0.173   & 0.095 & 0.045  & 0.045
					\\
					CP	      & 0.964 & 0.946  & 0.942  & 0.956   & 0.942  & 0.942  &0.958   &0.958   & 0.952 &0.948   &0.940 & 0.942 & 0.948 &0.940   & 0.956   
					\\
					\cdashline{1-16}
					Gaussian copula  & &  &   &   &    &   &  &  &  &  &
					\\
					\cdashline{2-16}
					Est.     &5.013   & 0.999 &2.007  &0.987   &-1.992  &-0.996 &-5.157 &-4.124  &-2.057 & 2.071 & 1.064 & 2.058  & 2.973  &0.594 &0.387
					\\
					SE       &0.287   & 0.028 & 0.031 &  0.315 & 0.362  & 0.407  & 0.491  & 0.443  & 0.261 & 0.280 &0.293 &0.183   & 0.095  & 0.040 &0.041
					\\
					SD       & 0.281  & 0.029 & 0.034 & 0.308 & 0.364  & 0.427 & 0.506  & 0.454  & 0.261 & 0.302  &0.312  & 0.184  & 0.109  & 0.054 & 0.050 
					\\
					RMSE  & 0.281   & 0.029 &0.034  & 0.308 & 0.363  & 0.426 &0.530  & 0.470 & 0.267  & 0.310  &0.318  & 0.193  & 0.112   & 0.055 & 0.051
					\\
					CP	    & 0.946   & 0.940 & 0.934 & 0.956 & 0.952  & 0.926 &0.944  &0.960 &0.950   & 0.936  &0.920  & 0.952 & 0.880  & 0.870 & 0.892
					\\
					\cdashline{1-16}
					$t$ copula ($df=5$)   & &  &   &   &    &   &  &  &  &  &
					\\
					\cdashline{2-16}
					Est.    & 5.008  & 0.999 & 2.001  &0.990   & -2.005 & -0.997 &-5.204 & -4.183  & -2.084 &2.095  &1.069  &2.081   &2.940    &0.608  &0.399
					\\
					SE      & 0.270  & 0.025 & 0.028  &  0.296 &  0.339 & 0.382  & 0.461  & 0.411    & 0.242   & 0.259 &0.272  & 0.172   & 0.101   & 0.043 &0.045
					\\
					SD      & 0.255  & 0.025 &  0.029 &  0.289 & 0.336  &  0.381 & 0.455  & 0.419   & 0.239   & 0.270 & 0.276  & 0.167  & 0.093  & 0.045 & 0.045
					\\
					RMSE & 0.255  & 0.025 & 0.029  & 0.289   & 0.335  & 0.381  & 0.498 & 0.457   & 0.253  & 0.286  &0.284   & 0.186  & 0.110   & 0.046 & 0.045
					\\
					CP	    & 0.962 & 0.948 & 0.940  & 0.960   & 0.948  & 0.946  &0.952  &0.958   & 0.950   & 0.940  &0.942  & 0.944  & 0.910   & 0.918  & 0.952  
					\\
					\cdashline{1-16}
					$t$ copula ($df=3$)  & &  &   &   &    &   &  &  &  &  &
					\\
					\cdashline{2-16}
					Est.    &5.017   &0.999   & 2.000   &0.989  & -2.006 & -0.997  &-5.013 & -4.010 & -2.005&2.014   &1.025   &2.000   &3.069   &0.600   &0.388
					\\
					SE      & 0.262  & 0.024  &  0.027 &  0.287 &  0.329 & 0.370    & 0.427 & 0.373  & 0.226  & 0.241  &0.255  & 0.159   & 0.103  & 0.044  & 0.047
					\\
					SD      & 0.257  & 0.025 &  0.029 &  0.294 & 0.341  &  0.378   & 0.436 & 0.401  & 0.229  & 0.259  &0.260  & 0.161   & 0.096  &  0.045 & 0.044 
					\\
					RMSE & 0.258  & 0.025 & 0.029  &0.294   &0.341   & 0.377    & 0.435  & 0.401 & 0.228  & 0.259  & 0.261  & 0.161   &0.118    &  0.045 & 0.046  
					\\
					CP	    & 0.954 & 0.934 & 0.932  & 0.950  & 0.934  & 0.934    &0.956  &0.940  &0.956   & 0.944  &0.942  & 0.948  & 0.928  & 0.950   & 0.960
					\\
					\hline
			\end{tabular}}
			\label{simt2}
	\end{table}
	\endgroup
	\vspace*{\fill}
\end{landscape}

\begin{landscape}
	\vspace*{\fill}
	\begingroup
	\setlength{\tabcolsep}{6pt} 
	\renewcommand{\arraystretch}{1.18} 
	\begin{table}[H]
			\tbl{\textbf{{\scriptsize Data are generated by the multivariate $t$ copula joint model with 4 degree of freedom and 80\% dropout (75\% censoring) rate and estimate the parameters by the multivariate Gaussian copula joint model and multivariate $t$ copula joint model with 3, 4 and 5 degree of freedom when the marginal distributions and $\bm R(\bm\rho)_{(t,\bm y)}$ are correctly specified.}}}
			{\begin{tabular}{lccccccccccccccccc}
					\toprule
					True value &$\beta_{01}$&$\beta_{11}$&$\beta_{21}$&$\beta_{31}$&$\beta_{41}$&$\beta_{51}$&$\beta_{12}$&$\beta_{22}$&$\beta_{32}$&$\beta_{42}$&$\beta_{52}$&$r$&$\sigma$&$\rho_{ty}$  &$\rho_{y}$
					\\
					& 5         & 1         & 2                 & 1             &  -2         &   -1        & -5        &-4      & -2       & 2   &1    & 2       &  3       &  0.6         & 0.4      
					\\
					\hline
					$t$ copula ($df=4$) &                   &                   &                    &                   &                   &                   &                    &                    &                     &                   &
					\\
					\cdashline{2-16}
					Est.       &5.017   & 0.998  & 2.002  & 0.985  &-2.004 & -1.008  & -5.169 & -4.159 &-2.071 &2.080  &1.069 & 2.066 & 2.983  & 0.607  & 0.393
					\\
					SE         & 0.281  & 0.031  &  0.036 &  0.310 & 0.355  & 0.401   & 0.495  & 0.482  & 0.274  & 0.289 &0.310 & 0.188  & 0.107  & 0.050  & 0.051
					\\
					SD         & 0.268 & 0.033  & 0.038  & 0.309  &0.358  &0.408    &0.512   &0.521   &0.285   &0.304  &0.322 &0.193   &0.099  & 0.054  &0.049  
					\\
					RMSE    & 0.268 & 0.033  & 0.038  &  0.309 &0.357   &0.408    &0.538   & 0.544  & 0.294 & 0.314  &0.329 &0.204   & 0.101 & 0.054  & 0.049
					\\
					CP	      & 0.962 & 0.942  & 0.932  & 0.952   & 0.946  & 0.948  &0.940   &0.956   & 0.934 &0.934   &0.946 & 0.942 & 0.956 &0.926   & 0.954   
					\\
					\cdashline{1-16}
					Gaussian copula  & &  &   &   &    &   &  &  &  &  &
					\\
					\cdashline{2-16}
					Est.     &5.011   & 0.999 &2.001   &0.982   &-1.989  &-0.999 &-5.180 &-4.156  &-2.071 & 2.085 & 1.069 & 2.070  & 2.976  &0.594 &0.386
					\\
					SE       &0.298  & 0.035 & 0.040 &  0.329 & 0.378  & 0.427  & 0.541  & 0.539  & 0.300 & 0.318 &0.338 &0.206   & 0.102  & 0.045 &0.046
					\\
					SD       & 0.290 & 0.038 & 0.044 & 0.331  & 0.385  & 0.445 & 0.559  & 0.561  & 0.312 & 0.341  &0.357  & 0.208  & 0.114  & 0.065 & 0.057 
					\\
					RMSE  & 0.290  & 0.038 &0.044  & 0.331 & 0.385  & 0.444 &0.586  & 0.581   & 0.319  & 0.351  &0.363  & 0.219  & 0.117  & 0.065 & 0.058
					\\
					CP	    & 0.956  & 0.922 & 0.932 & 0.946 & 0.942  & 0.942 &0.946  &0.960   &0.938   & 0.948  &0.926  & 0.950 & 0.884 & 0.844 & 0.864
					\\
					\cdashline{1-16}
					$t$ copula ($df=5$)   & &  &   &   &    &   &  &  &  &  &
					\\
					\cdashline{2-16}
					Est.    & 5.014  & 0.998 & 2.002  &0.988   & -2.003 & -1.009 &-5.220 & -4.202  & -2.093  &2.101  &1.080  &2.087    &2.950    &0.610  &0.399
					\\
					SE      & 0.284  & 0.032 & 0.037  &  0.313 &  0.359 & 0.406  & 0.508  & 0.499    & 0.280   & 0.296 &0.317  & 0.193   & 0.107   & 0.049 &0.051
					\\
					SD      & 0.268  & 0.033 &  0.039 &  0.308 & 0.361  &  0.411 & 0.520  & 0.523    & 0.289   & 0.309 & 0.326  & 0.196  & 0.100  & 0.054 & 0.050
					\\
					RMSE & 0.268  & 0.033 & 0.039  & 0.308   & 0.361  & 0.411  & 0.564 & 0.560    & 0.303   & 0.325  &0.335   & 0.214  & 0.111   & 0.054 & 0.050
					\\
					CP	    & 0.962 & 0.932 & 0.936  & 0.950   & 0.944  & 0.950  &0.944  &0.962   & 0.938   & 0.926  &0.946  & 0.940  & 0.922   & 0.904  & 0.950  
					\\
					\cdashline{1-16}
					$t$ copula ($df=3$)  & &  &   &   &    &   &  &  &  &  &
					\\
					\cdashline{2-16}
					Est.    &5.020   &0.998   & 2.002   &0.988  & -2.007 & -1.008  &-5.050 & -4.056 & -2.026&2.030   &1.044   &2.017   &3.057   &0.601   &0.386
					\\
					SE      & 0.277  & 0.031  &  0.035  &  0.305 &  0.350 & 0.396    & 0.472 & 0.453  & 0.263  & 0.278  &0.300  & 0.180   & 0.109  & 0.051  & 0.052
					\\
					SD      & 0.273  & 0.033 &  0.039  &  0.312 & 0.362  &  0.409   & 0.498 & 0.513  & 0.277   & 0.298  &0.315  & 0.188    & 0.101  &  0.053 & 0.048 
					\\
					RMSE & 0.273  & 0.033 & 0.039   &0.312   &0.362   & 0.409    & 0.500  & 0.515 & 0.278   & 0.299  & 0.318  & 0.189   &0.116    &  0.053 & 0.050  
					\\
					CP	    & 0.950 & 0.938 & 0.922   & 0.942  & 0.938  & 0.940    &0.942  &0.942  &0.942    & 0.934  &0.940  & 0.944   & 0.948  & 0.936   & 0.946
					\\
					\hline
			\end{tabular}}
			\label{simt3}
	\end{table}
	\endgroup
	\vspace*{\fill}
\end{landscape}

\end{document}